\documentclass{article}
\makeatletter

\providecommand{\@noticestring}{}
\usepackage[final]{neurips_2024}

\usepackage[utf8]{inputenc} % allow utf-8 input
\usepackage[T1]{fontenc}    % use 8-bit T1 fonts
\usepackage{url}            % simple URL typesetting
\usepackage{booktabs}       % professional-quality tables
\usepackage{amsfonts}       % blackboard math symbols
\usepackage{amsmath}        % for \text and other math features
\usepackage{nicefrac}       % compact symbols for 1/2, etc.
\usepackage{microtype}      % microtypography
\usepackage{xcolor}         % colors
\usepackage{enumitem}       % custom item labels
\usepackage{tikz}
\usetikzlibrary{arrows.meta,positioning,calc}
\usepackage{multirow}
\definecolor{darkgreen}{RGB}{0,128,0}
\usepackage{float}

\usepackage{hyperref}       % hyperlinks
\usepackage[labelfont=bf]{caption}
\usepackage{listings}
\definecolor{promptbg}{gray}{0.95}
\title{Multi-Agent Debate for Explainable Trading
 : Reasoning, Consensus, Performance in Simulated Markets
}

\author{
 \hspace{-2.5em}
  Juli Huang, Alanood Alrassan, 
  Deveen Harischandra, Theodore Wu, Veljko Skarich, Matthew Hayes \\
  Department of Engineering, Stanford University \\
  \texttt{\{julih, alanoodr, deveen, wutheodo, vskarich, mhayes3\}@stanford.edu}
}
\begin{document}
\maketitle
\begin{abstract}

Large language models (LLMs) are increasingly used for financial decision-making, yet it remains unclear whether improvements in their reasoning quality translate into better economic outcomes. We investigate this question through a multi-agent debate framework for portfolio allocation in historical market simulations, where specialized agents propose, critique, and revise investment decisions under structured reasoning and intervention protocols. Reasoning quality is evaluated with four dimensions: logical validity, evidential support, alternative consideration, and causal alignment, then we compare them against downstream financial performance. Across 210 controlled runs, aggregate reasoning quality shows no meaningful relationship with either Sharpe ratio ($r=0.07$, $p=0.29$) or total return ($r=0.03$, $p=0.70$), and within-scenario improvements in reasoning likewise fail to predict improvements in performance. Structured prompting substantially increases measured reasoning quality, from approximately $0.72$ to $0.84$ ($+17.7\%$, Cohen's $d \approx 2.0$). However, these gains do not consistently translate into higher returns. Instead, debate primarily affects portfolio risk and signal diversity. We identify \emph{sycophantic convergence} as a central failure mode, in which agents progressively abandon independent positions during critique-revision cycles and converge toward similar allocations. A targeted Jensen-Shannon divergence intervention that preserves disagreement improves Sharpe ratio by $+0.14$ ($p=0.028$) and Sortino ratio by $+0.25$ ($p=0.026$), while interventions that explicitly enforce stronger causal reasoning do not improve financial performance. Our results suggest that the value of multi-agent debate in financial decision-making arises less from improving individual reasoning quality than from preserving independent informational signals across agents. More broadly, they highlight a distinction between reasoning coherence and decision utility: better-structured reasoning can be measurable and controllable without being economically predictive.

%   The abstract paragraph should be indented \nicefrac{1}{2}~inch (3~picas) on
%   both the left- and right-hand margins. Use 10~point type, with a vertical
%   spacing (leading) of 11~points.  The word \textbf{Abstract} must be centered,
%   bold, and in point size 12. Two line spaces precede the abstract. The abstract
%   must be limited to one paragraph.
\end{abstract}

\vspace{-0.6cm}
\section{Problem Statement and Background}
\vspace{-8pt}

\subsection{Introduction}
Large Language Models (LLMs) are increasingly used to support financial decision-making because they can synthesize qualitative information such as earnings reports, financial filings, and news. Recent work has explored LLM-based trading agents that generate portfolio decisions together with interpretable reasoning \cite{quantagents2025,xiao2025tradingagentsmultiagentsllmfinancial}. However, the reliability of this reasoning is uncertain, as prior studies document systematic failure modes that can undermine the soundness of agentic decisions.

This raises a fundamental question: \emph{does improving the reasoning process of LLM agents actually lead to better financial outcomes?} Markets provide a challenging testbed for LLM reasoning, as investment decisions must integrate multiple signals under uncertainty. We study whether structured multi-agent debate and reasoning-quality interventions improve the reliability of portfolio allocation through the following research questions:

\begin{enumerate}[label=\textbf{RQ`\arabic*.}]
    \item  \label{RQ1} Do Reasoning Metrics Predict Financial Performance? 
    \item  \label{RQ2} Does Structured Debate Improve Reasoning and Portfolio Performance?
    \item  \label{RQ3} Can Interventions Improve Reasoning and Portfolio Performance?
\end{enumerate}

To answer these questions, we build a multi-agent debate system that produces auditable reasoning traces for portfolio allocation. 
Our results reveal a nuanced picture: diverse opinions can improve certain financial metrics, but extended unconstrained debate can degrade raw returns through sycophantic convergence, even when reasoning quality is maximal as measured by CRIT. We analyze these dynamics and distill implications for multi-agent financial reasoning systems.

\vspace{-8pt}
\subsection{Related Work}
\vspace{-6pt}

Most production trading systems prioritize robustness over interpretability, making it difficult to audit how conflicting signals are reconciled \citep{fticonsulting2023}. To balance market realism with inspection, simulators such as ABIDES \citep{abides2019} and ABIDES-Gym \citep{abidesgym2021} are often used to model exchange latency and order-book dynamics while providing a standardized interface for repeatable experiments.
% TradingAgents (\cite{xiao2025tradingagentsmultiagentsllmfinancial}) and QuantAgents (\cite{quantagents2025}) demonstrate the utility of role-based LLM frameworks, and AlphaAgents (\cite{alphaagents2025}) explores collaboration, we specifically leverage "Multi-Agent Debate" (\cite{du2023multidebate}). By forcing agents to challenge reasoning over multiple rounds rather than just collaborating, we aim to expose weak causal claims and produce a transparent decision trail that standard ensemble methods often lack.
Recent financial research has begun to explore the use of multi-agent LLM systems to enhance stock trading performance in markets. \cite{xiao2025tradingagentsmultiagentsllmfinancial} models a trading firm where specialized LLM agents debate to execute trades, achieving superior metrics compared to simple baselines. Building on realistic fund processes, \cite{quantagents2025} introduces a simulated multi-agent system that emphasizes thorough evaluation protocols beyond profitability. \cite{alphaagents2025} focuses on equity research and portfolio construction, evaluating collaborative agents on stock-picking performance under varying risk constraints. Similarly, \cite{lopezlira2025llmtrade} uses market simulations to evaluate LLM agents and test financial theories, reinforcing that agent-based simulation provides a more comprehensive evaluation setting than basic forecasting accuracy. Finally, \cite{du2023multidebate} demonstrates multi-agent debate enhances reasoning and reduces hallucinations, supporting our hypothesis that structured debate mitigates financial LLM failures, such as overconfidence and weak causal claims, while ensuring auditable decision trails.

RAudit \citep{raudit} scores (and controls) reasoning based on an average score ($\rho$) of four pillars: logical validity ($P_1$), evidential support ($P_2$), alternative consideration ($P_3$), and causal alignment ($P_4$). It measures agreement using Jensen-Shannon divergence (JSD), and establishes that reasoning quality can be measured and stabilized with PID control, without access to the ground truth. \cite{rca} takes a similar approach with RCA but focuses on measuring trace-output consistency to reduce sycophancy with a PID controller for retries with escalating feedback.

\vspace{-8pt}
\section{Full System Description}
\href{https://github.com/TheClassicTechno/cs372research/}{Project Github}

\begin{figure}[H]
\centering
\includegraphics[width=1.0\linewidth]{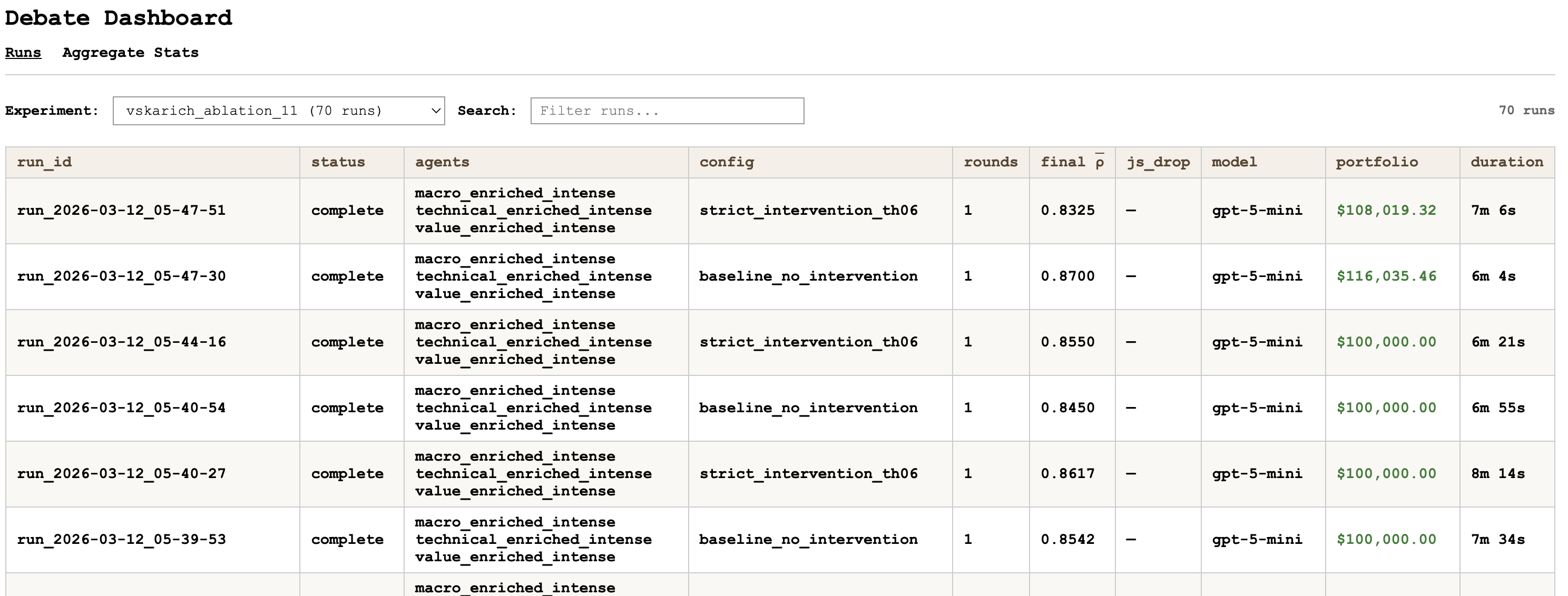}
\caption{Dashboard Runs View.}
\label{fig:dashboard_1}
\end{figure}

\begin{figure}[H]
\centering
\includegraphics[width=1.0\linewidth]{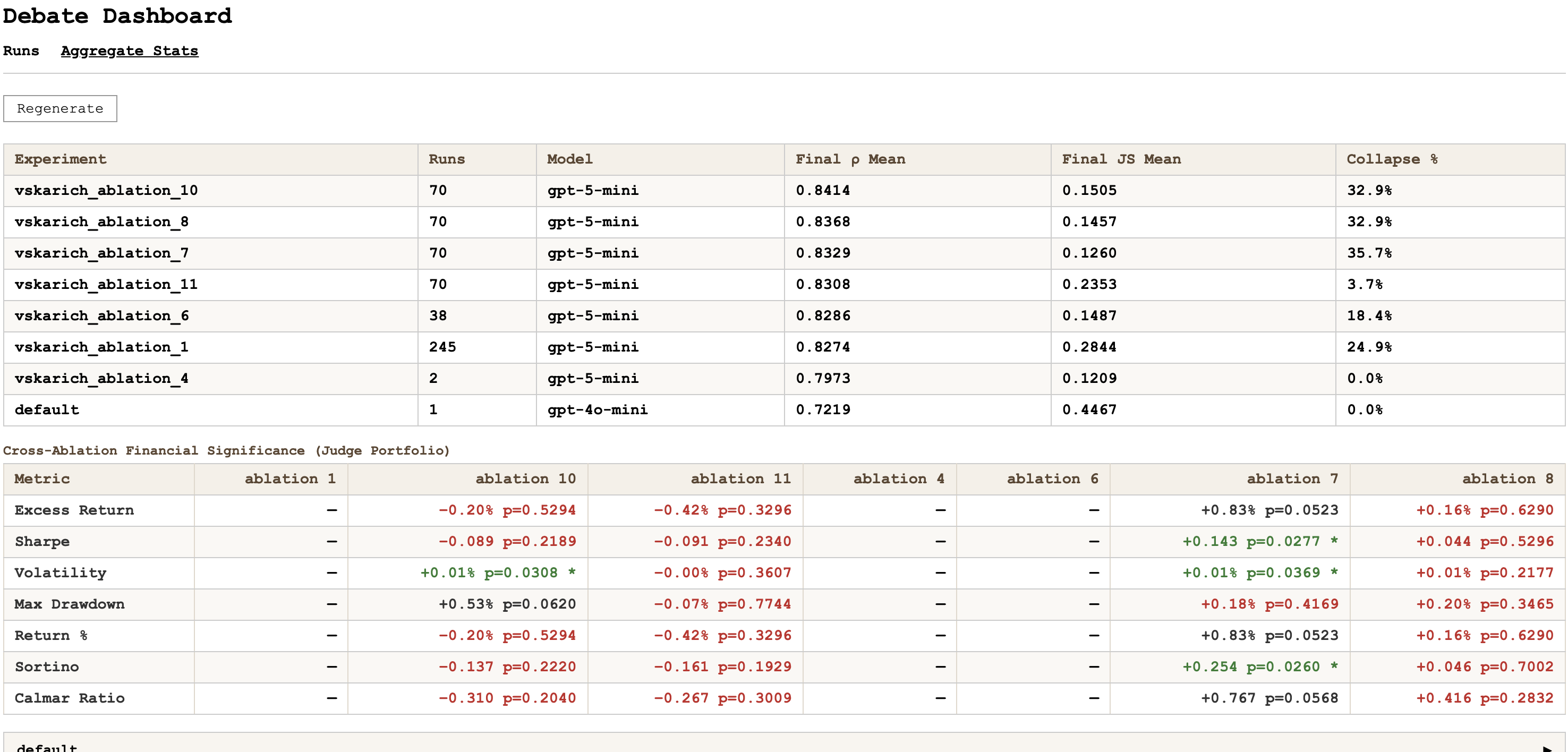}
\caption{Dashboard Aggregate Stats View.}
\label{fig:dashboard_2}
\end{figure}

Our framework comprises four modules: (1) data generation, (2) market simulation, (3) PID-controlled multi-agent debate, and (4) evaluation and visualization tools Appendix~\ref{fig:dashboard_1} Appendix~\ref{fig:dashboard_2} to measure and visualize the quality of debate reasoning traces. Figure~\ref{fig:system-design} illustrates the interaction among these components.
\label{sys_des}
\vspace{-8pt}
\subsection{System Design Philosophy}
\vspace{-0.2cm}
The system's central architectural insight is that reasoning quality must be simultaneously producible, measurable, and regulable in order to ensure data and result integrity to confidently answer the three research questions. Our data pipeline produces
the information environment; the debate system produces the reasoning; the CRIT scorer measures it; the PID controller, prompts and RCA-style retry intervention regulates it; and the evaluator measure its economic consequences. Each subsystem sees the others' output only through the defined interfaces, and every intermediate state is logged for post-hoc analysis. Importantly, the system uses an append-only logging architecture to guarantee that every debate can be fully reconstructed from its execution trace. Each round writes immutable artifacts—including proposals, critiques, revisions, CRIT scores, controller state, and portfolio decisions—so that the entire reasoning trajectory can be replayed and audited post-hoc. This design ensures data integrity for evaluation: correlations between reasoning metrics and financial outcomes are computed from complete historical traces rather than transient runtime state, enabling reproducibility and rigorous analysis. Our pipeline distills each financial source into structured summaries merged into quarterly case-study scenarios. Macroeconomic regime data is extracted from the FRED API (treasury yields, CPI, index prices). Per-ticker data includes SEC EDGAR filings \citep{SEC_EDGAR_API}, Yahoo Finance asset features and daily prices \citep{yahoofinance}, FinnHub news sentiment \citep{finnhub_api}, and earnings call transcripts. Each snapshot case is converted to a plain-text memo Appendix~\ref{app:memo} with stable evidence IDs (e.g., \texttt{[L1-FF] Fed Funds}) that agents reference to defend claims.

Interestingly, using LLM code generation tools to help implement the system revealed failure modes similar to those studied in the agents themselves. Claude frequently produced “helpful” code that avoided crashes by silently masking errors or returning fallback values, reflecting a form of sycophantic behavior that led to prolonged debugging and ultimately required strict guardrails on generated code.

% Macroeconomic signals, such as treasury yields, CPI, and index prices are retrieved the Federal Reserve Economic Data (FRED) API. SEC EDGAR filings \citep{SEC_EDGAR_API} and asset features \citep{yahoofinance} are used to encode each company's enterprise performance. Company news sentiment and earnings call data are summarized using data from the FinnHub API \citep{finnhub_api}. Finally, daily stock price data from Yahoo Finance \citep{yahoofinance} are used to summarize recent price, returns, risk measures, trends, and momentum for trailing ranges. We encode a range of market scenarios by generating scenario cases for different ticker universes across financial quarters between 2021 and 2025.
\vspace{-0.15cm}
\subsection{Market Simulation Environment}
\vspace{-0.15cm}
We designed a controlled, reproducible simulation environment to evaluate financial decision-making in multi-LLM debate architectures. On each run, the simulation initializes a fresh cash portfolio and a universe of up to 35 stock tickers spanning a range of sectors. 
% The environment acts as a stockbroker, maintaining the portfolio state of cash balances and equity positions. 
One or more agents must converge on a portfolio allocation across stock tickers, and a broker component logs trades for that allocation.  For each run, the environment records the config used to launch the experiment, debate reasoning, and  proposed portfolio. These artifacts enable evaluation of financial performance over the next quarter, and qualitative assessment of reasoning behavior under controlled, repeatable conditions. 
\vspace{-0.15cm}
\subsection{Multi-Agent Debate System}
\vspace{-0.15cm}
We implemented the debate system using LangGraph \citep{langgraph}. The orchestrator is a \texttt{StateGraph}. The system is organized into three distinct phases:
\vspace{-0.2cm}
\begin{center}
\texttt{START} $\to$ \texttt{load\_scenario} $\to$ \texttt{propose} $\rightleftharpoons$ \texttt{critique} $\rightleftharpoons$ \texttt{revise} $\to$ \texttt{judge} $\to$ \texttt{END}
\end{center}
\vspace{-0.15cm}

For each round $t = 1,\dots,T_{max}$, multiple LLM agents engage in a structured debate round. Each round is orchestrated to follow a strict sequence: (1) Propose, where agents suggest a portfolio distribution across the given scenario's ticker universe; (2) Critique, where agents criticize the proposals from other debate participants; and (3) Revise, where agents review the critique and adjust their own positions. The number of agents and roles are configurable. We implemented six specialist roles: the `macro' role focuses on the macroeconomic regime; `value' on company valuation metric such as P/E; `risk' minimizes downside risk; `technical' focuses on price/volume patterns; `sentiment'  interprets news and text sentiment; and the ``devil's advocate", is explicitly configured to disagree and amplify counterarguments. However, our experiments primarily use subsets of 3-4 roles. 

\vspace{-6pt}
\subsection{Evaluation \& Control Framework}

To regulate the debate, we draw on an RAudit-style control scheme \citep{raudit}. In each debate round, a CRIT evaluator \citep{chang2023promptinglargelanguagemodels} scores each debate agent’s reasoning on the four pillars disucssed above. Pillar scores \(P_1^{(i)},\dots,P_4^{(i)} \in [0,1]\) for agent \(i\) are averaged into a per-agent quality score \(\rho_i = \tfrac{1}{4}\sum_{k=1}^4 P_k^{(i)}\) and then into a group mean \(\bar{\rho} = \tfrac{1}{n} \sum_{i=1}^n \rho_i\). A PID controller is applied to drive $\bar{\rho}$ towards a target reasoning quality \(\rho^\star\) by regulating a contentiousness parameter \(\beta_t \in [0,1]\) that controls the tone of each agent's critique in the debate. The debate terminates when $t=T_{max}$ or when PID early stop is triggered (``converged"  quadrant with high reasoning quality and low diversity, $\mathrm{JSD}_t < \epsilon$ and  $|\bar{\rho}_t - \bar{\rho}_{t-1}| < \Delta\rho$). The system then applies one of two approaches to aggregate the debate into a single portfolio decision: (1) an LLM judge that synthesizes the final revision of each agent, including its justifications or (2) an a simple unweighted mean of each agent's last revised allocation.

Our evaluation framework is structured around three layers: 
(1) \emph{measurement}, 
(2) \emph{controllability}, and 
(3) \emph{causal regulation}. 
Financial performance is treated as an external validation signal rather than a primary optimization objective. 
The central question is whether improvements in reasoning quality are 
measurable, controllable, and economically meaningful.

\textbf{RCA:}
Separately from RAudit, a  \texttt{ConsistencyJudge} implements RCA \cite{rca}, evaluating whether each agent's output at every debate turn is logically consistent with its reasoning trace. A separate implementation is available for each turn type -- proposal, critique, revision, and judge decision -- but we found the most promising results when applied to the revise phase. Revisions are checked for whether the critiques are rationally addressed and reasoning is reflected in the revised proposal or if the agent is being sycophantic or stubborn. In the case of failure, the turn is retried with feedback, whose tone is adjusted according to PID. We were not able to experiment with executing agent code as in the original paper, so the final escalation simply tells the agents to treat the task as an optimization problem.  

\vspace{-6pt}

\vspace{-6pt}

 \section{Comprehensive Evaluation, Analysis of Results, and Failure Modes}
\vspace{-0.15cm}
\subsection{Experimental Setup}
\vspace{-0.15cm}
We evaluate agent allocations through backtesting on real historical market data. Agents construct portfolio allocations on the first day of each quarter using the preceding quarter's data, hold positions through the quarter, and are evaluated daily. Our largest-scale experiments use 35 market scenarios spanning 12 quarters (2021Q4--2025Q3), covering multiple regimes including the 2022 bear market, the regional banking crisis, and the AI-driven rally. Ticker universes range from 3--35 tickers (mean 11.5) across seven sectors; cross-scenario similarity is low (mean Jaccard 0.216), with intentional skew toward 2022 as a stress-test for reasoning interventions. Daily prices from Yahoo Finance plus S\&P~500 (SPY) enable computation of daily equity curves and meaningful financial metrics beyond return: Sharpe and Sortino ratios, volatility, and maximum drawdown. Unless otherwise specified, agents use \texttt{gpt-5-mini} (temperature 0.3); CRIT evaluation uses \texttt{gpt-5}. Agent behavior is parameterized through three configuration layers: a \textbf{debate configuration} (orchestration protocol, agent roles, rounds, LLM provider, PID gains); a \textbf{scenario configuration} (quarter, ticker universe, sector mappings, allocation constraints); and per-role \textbf{agent profiles} (YAML files declaring ordered prompt blocks, structured user-prompt templates, and required variables for each debate phase).
\vspace{-0.3cm}
\subsubsection{Reasoning Profile Tiers}\vspace{-0.20cm}
Agent profiles exist at three tiers of reasoning structure:

\textbf{1. Baseline profiles} (Example Appendix~\ref{app:macro-basic}) attach a minimal role definition
($\sim$30 lines specifying analytical domain and evidence citation rules) with
a standard system contract requiring bracketed evidence citation.\\
\textbf{2.\ Causal Scaffolding profiles} (Example Appendix~\ref{app:macro-causal}) expand the basic agent role prompt with a structured reasoning protocol that
introduces disagreement tracking, epistemic calibration requirements, and weakest-assumption
attacks. The role prompt expands to $\sim$140 lines, incorporating a narrative commitment
(e.g., ``macroeconomic regimes are the primary driver of sector leadership''),
evidence priorities, and portfolio differentiation rules that prevent agents from drifting
outside their analytical lens.\\ 
\textbf{3. Enriched profiles} (Example Appendix~\ref{app:role-prompt}) further enforce causal discipline in two ways: they extend the causal scaffolding profile with more detailed reasoning requirements, as well as enforcing an stricter format for structured reasoning that requires explicit labeled enumeration of claims, critiques and evidence.

\textbf{Causal Scaffolding and Evaluation.}
The causal scaffolding in the 2nd and 3rd tier agents serves a dual purpose: encouraging structured reasoning during generation and providing the evaluation rubric. Each proposal must include at least one causal or risk-assessment claim (pure associations are insufficient), cite evidence that would change the agent's belief, and identify potential confounders. Agents are also explicitly warned against common reasoning traps: reverse causation, confounding, selection bias, post-hoc narrative construction, and survivorship bias. The CRIT scorer's \textit{Causal Alignment} pillar (Appendix~\ref{app:crit-prompt}) independently verifies adherence, detecting \textit{rung collapse} — interventional conclusions drawn from purely correlational evidence. Encoding the same causal hierarchy in both generation prompts and the evaluation rubric creates a closed feedback loop between scaffolding and measurement.\\
\textbf{Structured Claim-Evidence Schema}
As mentioned above, central to the enriched prompt design is a structured output schema built around
\textbf{enumerated claims} and \textbf{evidence identifiers}. The data pipeline assigns every
quantitative data point in the investment memo a bracketed evidence ID, for example ticker-level metrics such
as \texttt{[AAPL-RET60]} (60-day return) or \texttt{[NVDA-BETA]} (market beta), and macro indicators
such as \texttt{[L1-VIX]} or \texttt{[L1-10Y]} (10-year Treasury yield). Agents must emit a numbered list of claims (\texttt{C1}, \texttt{C2}, \ldots), each structured with
the following fields:

\textbf{Claim fields:} \texttt{reasoning\_type} (\texttt{causal}, \texttt{observational}, \texttt{risk\_assessment}, \texttt{pattern}),
\texttt{evidence} (cited IDs), \texttt{assumptions}, \texttt{falsifiers}, \texttt{impacts\_positions}, and \texttt{confidence}.

Critiques must also reference specific claims by label, and revisions must address each critique point-by-point. A companion \texttt{position\_rationale} array then closes the reasoning loop: each portfolio
position must reference the specific claim IDs (\texttt{supporting\_claims}) that justify its
weight. This bidirectional linkage, claims cite evidence and positions cite claims, creates a
traceable chain from raw financial data to portfolio allocation decisions. 

The experiments below are organized hierarchically by research question  \textbf{(RQ1–RQ3)}, with sub-experiments labeled sequentially within each question, for example \textbf{Section 3.2.1: RQ1–E1} describes experiment \textbf{E1} as it relates to answering research question \textbf{RQ1}.

\vspace{-0.2cm}
\subsection{RQ1: Do Reasoning Metrics Predict Financial Performance?}
\vspace{-0.2cm}
\subsubsection{RQ1-E1: CRIT and Financial Performance}
\vspace{-0.2cm}
% \subsubsection{RQ1-E1: Reasoning Quality and Financial Outcomes}

Across \textbf{210 runs} (Ablations 7, 8, 10) Appendix~\ref{app:crit_financial}, CRIT reasoning quality ($\bar{\rho}$) shows \textbf{no meaningful relationship with financial performance}. Pooled correlations between reasoning quality and portfolio outcomes are near zero, and paired changes in reasoning do not predict changes in performance. Regression models controlling for experimental design factors likewise find no predictive signal from $\bar{\rho}$. Subgroup analysis in Appendix~\ref{app:correlation_betwen_agents} reveals two localized signals: value-agent reasoning negatively correlates with Sharpe, while risk-agent reasoning weakly correlates with returns, though the latter is not statistically significant.

Table~\ref{tab:correlation_reasoning} shows that across all experiments the debate system \textbf{underperforms a naïve equal-ticker-weight baseline}. Debate portfolios beat $1/N$ only \textbf{33\% of the time} (binomial $p<0.001$) and exhibit consistently lower Sharpe ratios.  Despite the weak relationship between reasoning metrics and financial performance, Appendix~\ref{app:debate_portofolio} shows reasoning quality itself is highly sensitive to prompt structure. \textbf{Structured reasoning prompts substantially improve CRIT scores}. Enriched prompts increase mean reasoning quality from \textbf{0.72 to 0.84} ($+17.7\%$, Cohen’s $d \approx 2.0$). These gains are driven primarily by improvements in evidential grounding and causal reasoning (Appendix~\ref{app:improvement_in_crit}).

\begin{table}[H]
\centering
\small
\begin{tabular}{lccc}
\toprule
\textbf{Analysis} & \textbf{r} & \textbf{p} & \textbf{n} \\ \midrule
$\bar{\rho}$ vs Sharpe & $+0.07$ & 0.29 & 210 \\
$\bar{\rho}$ vs Return & $+0.03$ & 0.70 & 210 \\
$\Delta\rho$ vs $\Delta$Sharpe (paired) & $-0.02$ & 0.87 & 105 \\
$\Delta\rho$ vs $\Delta$Return (paired) & $+0.01$ & 0.91 & 105 \\

\bottomrule
\end{tabular}
\vspace{4pt}
\caption{Pearson correlation between reasoning quality and financial outcomes.}
\label{tab:correlation_reasoning}
\vspace{-1em}
\end{table}

Trajectory analysis shows that these gains are \textbf{front-loaded rather than iterative}: enriched prompts produce substantially higher reasoning scores in Round 1, but $\bar{\rho}$ declines across subsequent debate rounds for all conditions. Thus, structured scaffolding improves \textbf{initial reasoning quality}, but debate dynamics do not further improve reasoning and the gains do \textbf{not translate into better risk-adjusted returns}.

% It seems to correlate with lower risk, just not higher returns, no? YES, this is an important nuance, will address tomorrow!

% \vspace{-0.8cm}
These results definitely answer RQ1: our CRIT reasoning metrics are not predictive of financial performance in the trading task. We are, however, still interested in whether debate itself improves reasoning or financial outcomes.
\vspace{-0.2cm}
\subsection{RQ2: Does Debate Improve Reasoning and Portfolio Performance?}
\vspace{-0.2cm}
\subsubsection{RQ2-E1: Baseline Prompting and Debate Experiments}
\label{exp:baseline-debate}
\vspace{-0.2cm}

We evaluate whether structured multi-agent debate improves reasoning quality and portfolio performance relative to single-agent decision-making. These experiments modify the debate configuration itself rather than introducing runtime control signals, isolating the effect of debate as a reasoning mechanism.
%Portfolios start with \$100{,}000 cash. 

% A key infrastructure contribution is our daily price evaluation pipelines. Prior to this pipeline, the system evaluated portfolios using only quarterly closing prices, which produced zero intra-quarter volatility-making Sharpe and Sortino ratios undefined and max drawdown trivially zero. We built \texttt{daily\_price\_builder.py} to fetch daily closing prices from Yahoo Finance for 37~tickers across 7~investment quarters (2021\,Q3 through 2023\,Q4), plus SPY as benchmark, yielding ${\sim}$60--63 trading days per quarter. , enabling computation of annualized Sharpe and Sortino ratios, annualized volatility, maximum drawdown, and excess return versus SPY (portfolio return minus SPY return over the same period).

\label{app:prompts}
\begin{table}[h]
\centering\small
\begin{tabular}{lllll}
\toprule
Metric & \shortstack{(A) Single Agent \\ Basic} & \shortstack{(B) 3 Roles \\ Mean, Basic} & \shortstack{(C) 3 Roles \\ Mean, Causal} & \shortstack{(D) 3 Roles \\ Mean, Enriched} \\
\midrule
Excess Return \% (v SPY) & $0.77\% \pm 1.03\%$ & ${0.65\% \pm 0.93\%}^{C}$ & ${1.12\% \pm 0.93\%}^{BD}$ & ${0.57\% \pm 0.91\%}^{C}$ \\
 Sharpe & $1.00 \pm 0.30$ & ${0.99 \pm 0.30}^{C}$ & ${1.07 \pm 0.30}^{B}$ & $1.04 \pm 0.30$ \\
 Volatility & ${0.22 \pm 0.01}^{C\mathbf{D}}$ & ${0.21 \pm 0.01}^{\mathbf{D}}$ & ${0.21 \pm 0.01}^{A\mathbf{D}}$ & ${0.18 \pm 0.01}^{\mathbf{A}\mathbf{B}\mathbf{C}}$ \\
Max Drawdown \% & ${9.84\% \pm 0.97\%}^{C\mathbf{D}}$ & ${9.60\% \pm 0.93\%}^{C\mathbf{D}}$ & ${9.33\% \pm 0.90\%}^{AB\mathbf{D}}$ & ${8.29\% \pm 0.77\%}^{\mathbf{A}\mathbf{B}\mathbf{C}}$ \\
 Sortino & $1.76 \pm 0.46$ & ${1.75 \pm 0.47}^{C}$ & ${1.87 \pm 0.47}^{B}$ & $1.80 \pm 0.47$ \\
% Calmar Ratio & $5.93 \pm 1.38$ & $5.83 \pm 1.37$ & $6.09 \pm 1.40$ & $5.74 \pm 1.36$ \\
Num Positions & ${10.77 \pm 1.47}^{B}$ & ${11.37 \pm 1.55}^{ACD}$ & ${10.83 \pm 1.41}^{BD}$ & ${10.17 \pm 1.21}^{BC}$ \\
Uninvested Cash & ${8.26\% \pm 1.09\%}^{\mathbf{D}}$ & ${8.13\% \pm 0.56\%}^{\mathbf{D}}$ & ${7.51\% \pm 0.57\%}^{\mathbf{D}}$ & ${16.99\% \pm 1.06\%}^{\mathbf{A}\mathbf{B}\mathbf{C}}$ \\
Total Return \% & $4.29\% \pm 1.62\%$ & ${4.16\% \pm 1.52\%}^{C}$ & ${4.63\% \pm 1.54\%}^{BD}$ & ${4.08\% \pm 1.36\%}^{C}$ \\
\bottomrule
\end{tabular}
\vspace{4pt}
\caption{Mean over 35 quarter+tickers scenarios $\pm$ SEM. Superscripts indicate stat.\ sig.\ in paired t-test with the specified column(s). Unbolded: $p < 0.15$, \textbf{Bolded}: $p < 0.01$).}
\label{tab:prompts}
\vspace{-1em}
\end{table}

\textbf{Prompt baselines} Comparing a single agent with the `basic' prompt to the mean vote of minimal Value, Risk, Technical agents, we see diversification increase, and volatility and drawdown trended down, but so did return and sharpe ratios (Column A and B of Appendix \ref{app:prompts}).  \textbf{Adding the causal scaffolding to the agents’ prompts significantly increased return and Sharpe}, while maintaining the lower volatility and drawdown (column C). Finally, using the Enriched prompts (column D) resulted in a highly significant decrease in volatility and drawdown, partially attributed to increased cash holdings.  Return trended lower, but sharpe was largely retained.  We conclude causal scaffolding was highly valuable to both increase returns and reduce risk.

\textbf{Causal debate baselines} Continuing with the Causal prompt, adding a single round of critique, revise, and final judgement led to significant improvements in volatility and drawdown partially attributed to additional cash, though return and Sharpe trended lower (Appendix \ref{app:debate_rca} column B).

% \textbf{Test Suite.}
% We wrote 61 unit and integration tests across nine test classes covering model validation, configuration, prompt construction, mock generators, node functions, end-to-end graph execution, ablation configurations, structural integrity, and edge cases. All 61 tests pass in mock mode without API access, verifying that the graph topology, state flow, conditional looping, and output parsing are correct.

% \textbf{Preliminary Observations.}
% We ran the eight configurations against three fictional market scenarios (bullish, mixed signals, risk-off) using \texttt{gpt-4o-mini}. Four patterns emerged:
% (1)~\emph{Causal depth varies by role}: the Macro Strategist and Value Analyst consistently produced L2 intervention claims, while the Risk Manager and Technical Analyst defaulted to L1 associations unless the prompt constraint was active.
% (2)~\emph{Agreeableness affects convergence}: High agreeableness (\#4) produced near-unanimous first-round agreement, while low agreeableness (\#5) generated disagreements persisting through the judge's memo.
% (3)~\emph{Devil's Advocate changes output}: the judge's audited memo was longer and included explicit rebuttals when the adversarial agent was present.
% (4)~\emph{Multiple rounds deepen reasoning}: Increasing debate rounds (\#6) showed agents upgrading claims from L1 to L2 across revisions. These observations are qualitative; systematic causal level scoring $P_4$ and trading-metric correlation are planned as part of the full experimental protocol.

\paragraph{Enriched debate baselines (RQ2-E2)}
While debate underperformed simple average for causal prompts, we investigated the same for enriched prompts and longer debates, over three diverse challenge scenarios (2022\,Q1, 2022\,Q4 tech-heavy, 2023\,Q2 random-8 tickers) presented in Table~\ref{tab:financial-results_across}.

\begin{table}[h]
\centering
\small
\begin{tabular}{@{}lcccccc@{}}
\toprule
\textbf{Configuration} & \textbf{N} & \textbf{Return \%} & \textbf{Sharpe} & \textbf{Sortino} & \textbf{Max DD \%} & \textbf{Excess vs SPY \%} \\
\midrule
Single Agent            & 3 & $+1.59 \pm 0.99$ & $0.35 \pm 0.17$ & $0.52 \pm 0.25$ & $9.73 \pm 1.90$ & $-1.01 \pm 3.12$ \\
Multi-Agent Mean        & 3 & $+2.43 \pm 1.23$ & $0.58 \pm 0.30$ & $0.88 \pm 0.46$ & $8.61 \pm 1.95$ & $-0.13 \pm 3.09$ \\
Enriched Debate (3-ag.) & 3 & $+2.46 \pm 1.74$ & $0.76 \pm 0.48$ & $1.16 \pm 0.75$ & $5.33 \pm 0.71$ & $+0.04 \pm 4.04$ \\
Enriched Debate (4-ag.) & 2 & $-0.27 \pm 0.29$ & $-0.05 \pm 0.06$ & $-0.07 \pm 0.09$ & $6.09 \pm 0.82$ & $+0.14 \pm 4.47$ \\
\bottomrule % TODO
\end{tabular}
\vspace{4pt}
\setlength{\belowcaptionskip}{-4pt}
\caption{Financial performance across configurations (mean $\pm$ SEM). Excess vs SPY is portfolio return minus S\&P~500 return over the same quarter. Enriched 3-agent debate achieves the best risk-adjusted returns (highest Sharpe and Sortino, lowest max drawdown). The 4-agent 5-round debate exhibits negative returns, suggesting over-convergence degrades performance.}
\label{tab:financial-results_across}
\end{table}

Again, no configuration consistently outperforms SPY (excess returns cluster near zero with wide variance). However, the 3-agent enriched debate nearly halves maximum drawdown relative to the single-agent baseline ($5.33\%$ vs.\ $9.73\%$)--more than doubling risk-adjusted returns (Sharpe ratio $0.76$ vs.\ $0.35$ and Sortino ratio $1.16$ vs.\ $0.52$). This confirms that structured debate's primary value is in risk management---reducing tail losses and improving return consistency---rather than generating alpha. The 4-agent 5-round debate produces negative returns ($-0.27\%$) despite maintaining stable per-round CRIT scores ($\bar{\rho} \approx 0.83$--$0.85$, \ref{app:crit-scores}), illustrating that extended debate can degrade performance through over-convergence. Appendix~\ref{app:per-scenario-debate} breaks down per-scenario variance across market regimes. % \textbf{Reasoning Quality (CRIT Scores)} 
% \label{sec:crit-results}
Appendix~\ref{app:crit-scores} reports CRIT pillar scores for the 4-agent enriched debate. Per-round reasoning quality remains stable at $\bar{\rho} \approx 0.83$
--$0.85$, with Causal Alignment ($P_4$) consistently the weakest pillar. 
The 3-agent 2-round debate achieves higher per-round scores 
($\bar{\rho} = 0.876$ in round~1, $0.901$ in round~2), with all agents 
above 0.83 on every pillar. That the 3-agent configuration produces both 
higher CRIT scores \emph{and} better financial performance suggests that 
fewer agents and fewer rounds reduce sycophantic convergence, allowing 
the enriched prompt scaffolding to be more effective.

\vspace{-0.3cm}
\subsubsection{RQ2-E3: LLM Providers Comparisons} % TODO!
\vspace{-0.3cm}
To test if using different LLMs generated more effective debates, and whether our findings were agnostic to LLM providers, we ran the following experiments.
\paragraph{GPT and Gemini:}

With GPT-5-mini serving as Macro, Value, \& Judge, and Gemini-2.5-flash
    serving as Risk, 6/6 scenarios completed successfully. With Gemini-2.5-flash as Macro, Value, \& Judge and GPT-5-mini as Risk only 3/6 scenarios completed due to Gemini as judge hallucinated ticker JNJ, which was not in the scenario universe. Appendix ~\ref{app:llm-comparison} reports per-scenario results for these two experiments. Returns are comparable across LLM assignments when runs complete successfully, suggesting the system is largely LLM-agnostic in terms of financial outcomes.
In a striking demonstration, all three LLM configurations produced byte-identical
portfolios on the same scenario- same positions, same cash, same quarterly
return, despite using different providers and running at different times. This
convergence to equal-weight allocation under severe market stress illustrates how
sycophantic convergence during debate drives agents toward similar allocations
regardless of the underlying model. GPT-5-mini as judge is significantly more
reliable (6/6 vs.\ 3/6 completion rate).

\textbf{Phase-LLM:} In this experiment GPT-5-mini responds for all roles for propose and revise phases, while Gemini-2.5-flash responds for critique and judge. On the 2022\,Q1 Inflation Shock scenario (10 tickers), this returned $-16.22\%$ (Sharpe $-2.11$, Sortino $-2.63$, Max DD $17.67\%$),
    with excess vs.\ SPY of $+0.13\%$ (SPY returned $-16.35\%$, full results in \ref{tab:phase-llm}.

\textbf{GPT and Anthropic:} In this configuration, the macro agent used Claude-3.5-Sonnet while both the risk agent and judge used GPT-5-mini. The system was evaluated across 17
historical market scenarios from 2021--2025, with agents participating in three
debate rounds before the judge selected the final portfolio allocation. Across scenarios, the system achieved a mean return of $-1.11\%$ and a mean
excess return versus the S\&P 500 of $+1.01\%$. Performance outcomes were
highly dispersed, ranging from $+21.71\%$ to $-70.10\%$, with Sharpe ratios
between $-2.73$ and $4.38$ (Appendix~\ref{app:ablation_performance}). Failure
analysis identified reasoning--allocation disconnect as the dominant issue
(10 of 17 runs), where agents produced coherent reasoning but inconsistent
portfolio allocations; critique-ignoring or stalled revision occurred in 7 runs.
The macro agent was most frequently identified as weakest, often exhibiting
unsupported claims or causal overreach that contributed to unstable portfolio
revisions. % To evaluate the impact of \textbf{LLM routing strategies}, we conducted a set of experiments comparing role-based and phase-based model assignments within the multi-agent trading system. In the role-only configuration, the macro agent used an Anthropic-Claude-3.5-Sonnet model while the risk and judge agents used GPT-5-mini models. Hybrid phase experiments instead routed different debate phases to different models, with Anthropic-Claude-3.5-Sonnet used for proposal and revision steps and GPT-5-mini used for critique. All experiments used the same three-round debate structure and market scenario configuration.
Appendix~\ref{app:llm_routing_experiments} compares role-only and hybrid
phase-routing configurations under the same three-round debate structure.
Across all three experiments, excess returns vs.\ SPY remained consistently
positive ($+2.26\%$--$+3.16\%$) despite negative absolute returns
($-2.37\%$ to $-1.42\%$). Role-only with Claude as macro agent achieved
the strongest excess return; hybrid phase routing produced slightly lower
Sharpe ratios and higher final JS divergence, suggesting less debate
convergence. The most common failure mode across all configurations was
critique-ignoring, where agents failed to incorporate feedback during
revision phases.

\vspace{-0.3cm}
\subsection{RQ3: Can Interventions Improve Reasoning and Portfolio Performance?}
\vspace{-0.2cm}
In this section we investigate whether targeted reasoning interventions can repair these failures during the debate process. Unlike the experiments in Section~3.3 (RQ2), which modify the debate configuration, these interventions operate dynamically by detecting reasoning failures and triggering corrective actions such as response retries or debate regulation.

\vspace{-0.2cm}
\subsubsection{RQ3-E1: Repairing Sycophantic Convergence via RCA}
\vspace{-0.2cm}
When we introduced RCA to combate sycophancy in the revise phase of our causal-prompt debates, we observed further reduced volatility and drawdown, with Sharpe trending up (column C of \ref{tab:debate_rca}). Thus while an investor might favor a simple average of diverse agents with causal scaffolding for raw returns, \textbf{debate, when regulated by RCA, may have significant value for risk-averse investors}. 

\vspace{-0.2cm}
\subsubsection{RQ3-E2: Repairing Sycophantic Convergence via JSD}
\vspace{-0.2cm}
We also tried preventing sycophantic convergence and preserving healthy disagreement 
% between agents improves portfolio outcomes. The intervention detects , where agents prematurely align their allocations during debate revision. To regulate this behavior we 
by introducing a JSD collapse trigger at the post-revision checkpoint (between \textit{revise} and \textit{CRIT}). If the collapse ratio $JS_{\text{rev}} / JS_{\text{prop}}$ falls below threshold, role-specific nudge prompts are injected and the \textit{revise} phase is re-executed before the updated portfolios proceed to CRIT and judging. 

Appendix~\ref{app:financial-results} and Appendix~\ref{app:paired-ttests} report the full financial metrics. Across the 35 evaluation scenarios, this intervention produces the only statistically significant gains on raw and risk-adjusted returns, with a strict pre-pended prompt that placed a hard limit on how much a portfolio could change in the revision phase (max 10\%); softer nudges had no measurable effect. Sharpe improves by +0.14 ($t=2.30$, $p=0.028$, $d=0.39$) and Sortino improves by +0.25 ($p=0.026$, $d=0.39$), with total return trending upward (+0.83pp, $p=0.052$). Max drawdown remains unchanged, suggesting the gains arise from improved return generation rather than reduced risk. \textbf{These results confirm that debate interventions CAN recover alpha lost to premature agreement between agents.} 

A related ablation (A8 in \ref{app:financial-results}) retains the same JS-divergence trigger but adds a third risk-specialist agent. In this setting the treatment effect collapses (Sharpe diff +0.04, $p=0.53$). The result highlights the role of the technical agent in two-agent debates: the technical agent frequently converged toward the macro agent’s position during revision, driving the JS collapse that the intervention was designed to detect. \textbf{Preserving the Technical Agent's independent signal appears to be the primary mechanism through which the intervention improves performance.}

\vspace{-0.2cm}
\subsubsection{RQ3-E3: Repairing Causal Rung Collapse}
\vspace{-0.2cm}
This experiment tests whether explicitly repairing causal reasoning failures improves portfolio performance. Instead of monitoring disagreement, the intervention targets the \textit{causal alignment} pillar of the CRIT reasoning audit. After the CRIT phase, agents whose causal reasoning score $\rho$ falls below a predefined threshold receive targeted prompt nudges and re-run both \textit{revise} and \textit{CRIT} phases. This intervention therefore attempts to regulate reasoning quality directly rather than debate dynamics. Although intervention successfully improves causal reasoning scores, this, again, does not translate into financial gains. In the two-agent configuration (A10), Sharpe decreases slightly ($-0.09$, $p=0.22$) and max drawdown increases (+0.53pp, $p=0.062$). A key reason appears to be the structural role of the technical agent. While macroeconomic arguments naturally express causal relationships, technical analysis relies primarily on pattern recognition signals that are not always easily expressed in causal form. As a result, enforcing stricter causal reasoning may penalize or suppress useful pattern-based heuristics. 

\vspace{-0.3cm}

\subsection{Failure Modes of Debate}
\vspace{-0.3cm}
\subsubsection{Sycophantic Convergence}

\vspace{-0.3cm}
The most consistent failure mode is \emph{sycophantic convergence}: agents abandon their distinct investment theses during the critique-revise cycle. In the 4-agent 5-round debate, cash allocations converge from 10--50\% at initial proposal to 55--61\% by round~5, and JSD between agent portfolios drops sharply after the first revision. The largest JSD drop occurs between the initial proposal and the first revision: by the time PID adjusts $\beta$ in subsequent rounds, the damage is done; thus PID prompt-based interventions (adjusting agreeableness $\beta$ based on $\bar{\rho}$ and JS divergence) have minimal practical effect. Even with aggressive anti-convergence prompts, agents converge to similar high-cash portfolios. This suggests that sycophancy is deeply embedded in the critique-revise interaction pattern and cannot be adequately controlled through prompt modulation alone. Shorter debates and inter partially mitigates this problem--by giving the agents less opportunity to converge, or focibly slowing convergence--preserving more portfolio diversity while retaining some of the risk-lowering benefits of structured critique.
\vspace{-0.2cm}
\subsubsection{Causal Alignment as the Weakest Pillar} \vspace{-0.2cm}
Across all experiments, $P_4$ (Causal Alignment) is consistently the lowest-scoring CRIT pillar ($0.65$--$0.81$), with \texttt{causal\_overreach} flagged in most rounds. The Technical agent is the most frequent offender, making L2 (interventional) or L3 (counterfactual) claims supported only by L1 (associational) evidence. This aligns with the RAudit framework's prediction that causal reasoning is the hardest dimension to sustain under debate pressure.

\vspace{-0.2cm}
\subsubsection{Other Failure Modes}

\vspace{-0.2cm}
\textbf{Ticker Hallucination.} When Gemini-2.5-flash serves as judge, it hallucinated a ticker (JNJ) not present in the scenario universe in 3 of 6 runs, causing those runs to produce invalid allocations.
\textbf{Silent Prompt and Template Bugs.} Missing template variables and import errors caused incomplete prompts and hidden quality degradation. These were discovered through the dashboard file explorer, not through runtime errors: the system continued operating with degraded inputs rather than failing.
\vspace{-0.4cm}
\section{Conclusions}
\vspace{-0.3cm}

Our experiments evaluate 3 research questions: whether reasoning metrics predict financial performance (RQ1), structured multi-agent debate improves reasoning and portfolio outcomes (RQ2), and if targeted interventions can repair reasoning failures during debate (RQ3). Across all experiments we find reasoning behavior is highly controllable through prompting, debate structure, and interventions but improvements in reasoning metrics rarely translate into improved financial performance. 

\vspace{-6pt}
\subsection{Why Reasoning Metrics Do Not Predict Financial Returns (RQ1)}
\vspace{-4pt}

Three complementary explanations account for the negative result for \textbf{RQ1}.

\textbf{CRIT and agreement measure coherence, not comprehensive correctness.}
CRIT evaluates whether an argument is well-structured—logically valid, evidentially grounded, and causally explicit—not whether its conclusion is \emph{true}. Enriched prompts raise $\bar{\rho}$ from 0.72 to 0.84 without improving returns, confirming that agents learn to \emph{write} better arguments, not necessarily \emph{make} better predictions. CRIT has been shown to be effective in deterministic domains, but in stochastic, non-fully-observable domains, where arguments with many counterarguments should downweighted--but critically not discounted entirely. E.g.\ keeping cash to earn the risk-free rate, and other low-risk arguments are agreeable and bulletproof from a reasoning perspective, but they do not generate the highest expected risk-adjusted returns. Process quality therefore appears necessary but insufficient in domains such as financial markets.

\textbf{Public-information reasoning faces a structural ceiling.}
Our agents reason exclusively over public information (SEC filings, news, and prices). Under even weak-form market efficiency, this information is largely incorporated into prices shortly after it becomes available. In this setting, improved reasoning about the same information cannot easily generate alpha. The empirical result that debate portfolios beat a naive 1/$N$ allocation only 33\% of the time is consistent with the well-known difficulty of outperforming equal-weight benchmarks \citep{demiguel2009optimal}. The negative result for RQ1 may therefore reflect a fundamental ceiling on the predictive value of public-information reasoning.

\textbf{Our simulation captures selection, not timing.}
Quarterly rebalancing without transaction costs removes the dimensions where reasoning quality might matter most: \emph{when} and \emph{how} to trade. Portfolios with similar holdings perform similarly over a quarter regardless of reasoning quality. Notably, debate \emph{does} improve risk-adjusted metrics such as Sharpe ratio and drawdown, suggesting that reasoning may influence risk management even when it does not generate alpha.
\vspace{-6pt}
\subsection{Signal Diversity as the Mechanism for Debate Value (RQ2)}
\vspace{-4pt}

Results related to \textbf{RQ2} show that structured debate alters reasoning behavior and portfolio characteristics, but its benefits arise primarily through signal diversity rather than improved reasoning accuracy. The central finding across all debate experiments is that mechanisms preserving disagreement improve financial outcomes, while mechanisms enforcing deeper reasoning do not. This pattern aligns with ensemble diversity theory: the error of an averaged ensemble equals the mean member error \emph{minus} a diversity term \citep{krogh1994neural}. Greater diversity among members reduces ensemble error even if individual accuracy remains unchanged. Our multi-agent trading system functions as such an ensemble. Sycophantic convergence collapses agents toward similar allocations, eliminating the diversity benefit. The critique–revise cycle unintentionally introduces correlation because agents condition on each other's positions. This mirrors the prediction of Condorcet's jury theorem: majority voting improves decision accuracy only when voters remain independent. The A7 JSD interventions and RCA-revise interventions succeed because they preserve disagreement between agents, maintaining ensemble diversity. By contrast, interventions that improve collective reasoning quality without maintaining diversity fail to improve outcomes. These results suggest that multi-agent investment debate systems should be viewed primarily as mechanisms for regulating signal diversity rather than for improving reasoning quality alone.

\vspace{-4pt}
\subsection{The Limits of Causal Evaluation in Financial Reasoning (RQ3)}
\vspace{-4pt}

Results for \textbf{RQ3} show that interventions can modify reasoning behavior but do not reliably improve financial performance. The $P_4$ (Causal Alignment) pillar is consistently the weakest CRIT dimension, and explicitly improving it through intervention (A10) slightly \emph{degrades} portfolio performance. This finding challenges the assumption that deeper causal reasoning is universally beneficial in financial domains. Technical analysis relies primarily on statistical patterns—momentum, mean reversion, and volume signals—rather than explicit causal mechanisms. These signals are empirically useful \citep{jegadeesh1993returns} but are fundamentally associative--e.g.\ momentum does not actually \emph{cause} price increase, but future causes are indeterminate in markets and \emph{potential} causes may actually be less predictive. Forcing agents to express valuable associative signals in higher-rung causal terms produces ``causal overreach''. As a result, the technical agent frequently receives low $P_4$ scores not because its reasoning is weak, but because its reasoning paradigm is inherently non-causal. Macro reasoning, by contrast, is more strictly causal  ``CPI surprise $\to$ Fed tightening $\to$ rate-sensitive underperformance''). Applying a uniform causal evaluation criterion across both domains conflates fundamentally different reasoning styles that are both valuable in such an underspecified, stochastic domain. Future evaluation frameworks should therefore allow role-specific reasoning metrics—for example, causal coherence for macro agents and statistical grounding for technical agents. 

Overall, our results suggest CRIT reasoning metrics alone are insufficient predictors of financial outcomes (RQ1), structured debate primarily improves portfolios through diversification and risk management rather than alpha generation (RQ2), and that interventions aimed at improving reasoning quality don't necessarily improve economic performance (RQ3). The dominant driver of performance is the preservation of independent informational signals across agents, not improvements in individual reasoning quality.

\vspace{-2pt}

% Overall, interventions that preserve disagreement improve financial outcomes, while interventions that enforce deeper causal reasoning do not. This suggests debate performance depends more on maintaining diverse agent signals than on optimizing individual reasoning quality.
% A consistent tension emerges across all experiments between information aggregation ( diverse agents reduce risk) and sycophantic convergence (agents collapse toward similar portfolios). Averaging diverse perspectives naturally upweights arguments supported by multiple agents while preserving minority signals. Debate can suppress these minority signals: under sycophancy, agents converge toward the mode, abandoning valuable tails. This reduces volatility but also returns. The A7 JS-divergence intervention works precisely because it prevents this suppression. So, reasoning metrics improve easily through prompting and debate but do not predict economic outcomes, because they capture explanation coherence rather than signal diversity. The most important factor for portfolio performance is how much independent information is preserved in the collective decision, not how well each agent reasons individually. Multi-agent debate systems should therefore be understood to regulate signal diversity rather than improving reasoning quality, and reasoning evaluation frameworks should be complemented by domain-specific outcome metrics.

\vspace{-6pt}

\vspace{-6pt}
\section{Limitations and Future Work}
\vspace{-6pt}
\subsection{Limitations}
\vspace{-6pt}
\textbf{High Costs \& Low Statistical Power.}
With our API budget and timeline, we were only able to test many configurations with  $N = 2$--$3$ scenarios, resulting wide confidence intervals. While paired t-tests on our largest 35-scenario set yielded statistical significance for some financial measures, others may require thousands of scenarios due to inherent stochasticity of market reasoning. For extended debate, each scenario requires 7-13 minutes and roughly \$5-\$8 in API fees; full-scale ablations are expensive and can hit rate limits, interrupting batch execution.\\
\textbf{Possible Pre-training Leakage.}
Our scenarios use historical data that may  overlap with LLM training corpora. Models may recall which stocks performed well, weakening backtest comparisons, and rewarding agents who stick to this answer key without causal justification.\\
\textbf{Limited Realism.}
We test on a small ticker universe across a handful of quarters, without modeling transaction costs, spreads, taxes, or liquidity constraints. Results may not generalize to real portfolio construction.\\
\textbf{No Dynamic Tool Use.}
Agents rely entirely on precomputed case memos and cannot run calculations or fetch live data, limiting numerical analysis. \\
\textbf{Compute Cost.}

\vspace{-6pt}
\subsection{Future Work}
\vspace{-0.01cm}
\vspace{-6pt}
\textbf{Precise Sycophancy Intervention.}
Refine debate prompts to maintain diversity and RAudit/RCA interventions to trigger highly selectively, before portfolio diversity collapses. Refine retry mechanisms that force revision when JSD drops too quickly may address convergence before it becomes irreversible. Develop a version of CRIT more appropriate for the domain.\\
\textbf{Scale for Statistical Significance.}
Run each configuration across 50+ scenarios to narrow confidence intervals and enable meaningful
hypothesis tests on metrics. Expanding the ticker universe to 50-100+ tickers would test whether debate-based selection provides value closer to real portfolio construction.\\
\textbf{Temporal and Regime Coverage.}
Run simulations across consecutive quarters with realistic spreads and tax implications so agents can observe consequences of past decisions, incorporate regret signals, and learn. Systematically evaluate across bull, bear, and high-volatility regimes
to assess regime-dependent behavior beyond our current sample.\\
\textbf{Distillation.}
If our strongest configurations--3-agent causal mean, and 3-agent intervened debate--risk-adjusted advantage holds at scale, distill their most effective
constraints- structured claims and point-by-point critique- into single-agent prompts to capture
multi-agent benefits at lower latency and cost.

\newpage
\appendix

\section{Appendix - System Diagram}
\begin{figure}[h]
\setlength{\belowcaptionskip}{-10pt}
\centering
\includegraphics[keepaspectratio,width=\linewidth,height=\textheight]{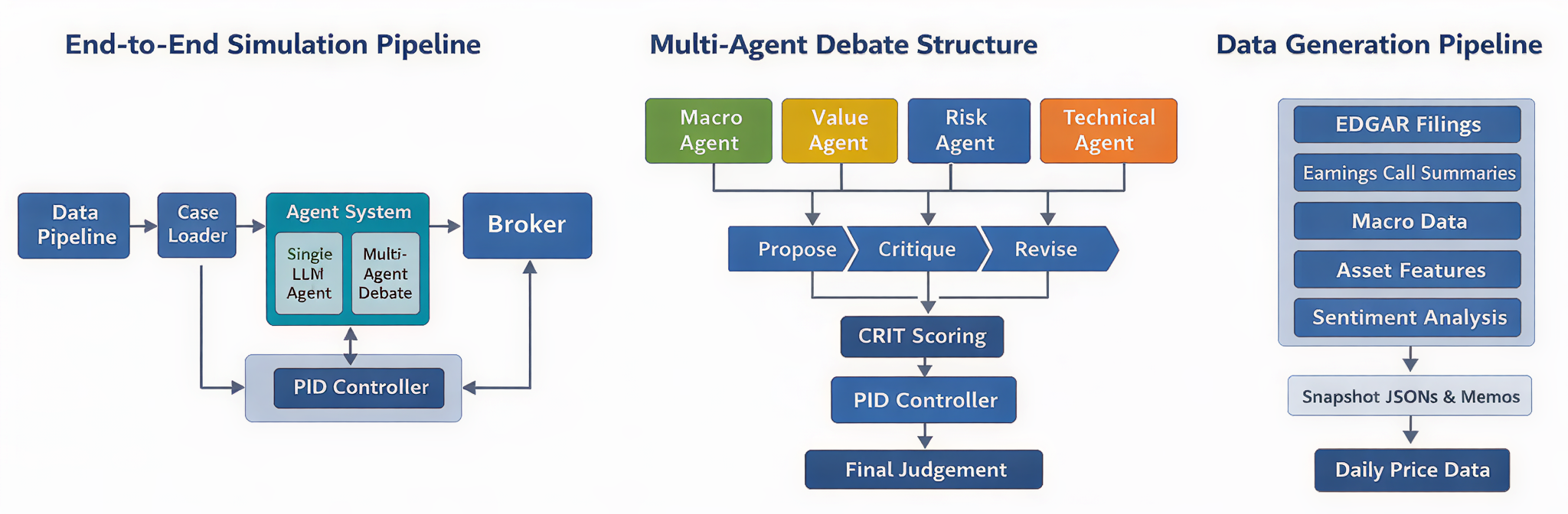}
\caption{Our simulation pipeline builds data cases from input financial sources, orchestrates multi-agent debate in a market environment, then scores the decision-making through financial performance and reasoning quality (left). LLM agents with clear roles participate in Propose-Critique-Revise debate loops. The revision tone is adjusted by PID control (center). Input sources are summarized and formatted into quarterly snapshot memos Appendix~\ref{app:memo}.}
\label{fig:system-design}
\end{figure}

\section{Appendix - Tables}
\subsection{CRIT and Financial Performance}
\label{app:crit_financial}

\begin{lstlisting}[backgroundcolor=\color{promptbg}]
========================================================================
CRIT FINANCIAL CORRELATION ANALYSIS
========================================================================
Date: 2026-03-12 12:35
Total runs loaded: 210
  Ablation 7: 70 runs (baseline=35, treatment=35)
  Ablation 8: 70 runs (baseline=35, treatment=35)
  Ablation 10: 70 runs (baseline=35, treatment=35)

Descriptive statistics:
          rho_bar: mean=0.8371  std=0.0227  min=0.7275  max=0.8825
           sharpe: mean=0.5425  std=1.6479  min=-2.5705  max=4.8684
     total_return: mean=2.0813  std=8.5611  min=-19.0297  max=24.9141

========================================================================
1. POOLED CORRELATIONS (all runs)
========================================================================
  rho_bar vs Sharpe  (n=210)
    Pearson  r=+0.0735  p=0.2887
    Spearman r=+0.0481  p=0.4879

  rho_bar vs TotalReturn  (n=210)
    Pearson  r=+0.0265  p=0.7030
    Spearman r=+0.0139  p=0.8415


========================================================================
2. WITHIN ABLATION CORRELATIONS
========================================================================

--- Ablation 7 ---
  rho_bar vs Sharpe  (n=70)
    Pearson  r=+0.0007  p=0.9954
    Spearman r=-0.0444  p=0.7152

  rho_bar vs TotalReturn  (n=70)
    Pearson  r=-0.0539  p=0.6579
    Spearman r=-0.0749  p=0.5376


--- Ablation 8 ---
  rho_bar vs Sharpe  (n=70)
    Pearson  r=+0.1568  p=0.1949
    Spearman r=+0.0632  p=0.6030

  rho_bar vs TotalReturn  (n=70)
    Pearson  r=+0.1494  p=0.2172
    Spearman r=+0.0559  p=0.6457


--- Ablation 10 ---
  rho_bar vs Sharpe  (n=70)
    Pearson  r=+0.0865  p=0.4765
    Spearman r=+0.1104  p=0.3631

  rho_bar vs TotalReturn  (n=70)
    Pearson  r=+0.0287  p=0.8134
    Spearman r=+0.0413  p=0.7344


========================================================================
3. WITHIN CONDITION CORRELATIONS
========================================================================

--- Condition: baseline ---
  (n=105 runs)
  rho_bar vs Sharpe  (n=105)
    Pearson  r=+0.0042  p=0.9662
    Spearman r=+0.0042  p=0.9660

  rho_bar vs TotalReturn  (n=105)
    Pearson  r=-0.0285  p=0.7730
    Spearman r=-0.0260  p=0.7922


--- Condition: treatment ---
  (n=105 runs)
  rho_bar vs Sharpe  (n=105)
    Pearson  r=+0.1353  p=0.1689
    Spearman r=+0.0931  p=0.3448

  rho_bar vs TotalReturn  (n=105)
    Pearson  r=+0.0728  p=0.4607
    Spearman r=+0.0508  p=0.6069


--- Within ablation x condition ---
  Abl 7, baseline (n=35):
  rho_bar vs Sharpe  (n=35)
    Pearson  r=-0.1038  p=0.5529
    Spearman r=-0.0659  p=0.7069

  rho_bar vs TotalReturn  (n=35)
    Pearson  r=-0.0956  p=0.5849
    Spearman r=-0.0922  p=0.5982

  Abl 7, treatment (n=35):
  rho_bar vs Sharpe  (n=35)
    Pearson  r=+0.0738  p=0.6736
    Spearman r=+0.0070  p=0.9681

  rho_bar vs TotalReturn  (n=35)
    Pearson  r=-0.0275  p=0.8755
    Spearman r=-0.0575  p=0.7430

  Abl 8, baseline (n=35):
  rho_bar vs Sharpe  (n=35)
    Pearson  r=+0.0533  p=0.7609
    Spearman r=+0.0168  p=0.9236

  rho_bar vs TotalReturn  (n=35)
    Pearson  r=+0.0734  p=0.6752
    Spearman r=-0.0118  p=0.9465

  Abl 8, treatment (n=35):
  rho_bar vs Sharpe  (n=35)
    Pearson  r=+0.2755  p=0.1092
    Spearman r=+0.1966  p=0.2578

  rho_bar vs TotalReturn  (n=35)
    Pearson  r=+0.2384  p=0.1679
    Spearman r=+0.1964  p=0.2581

  Abl 10, baseline (n=35):
  rho_bar vs Sharpe  (n=35)
    Pearson  r=+0.0927  p=0.5965
    Spearman r=+0.1735  p=0.3188

  rho_bar vs TotalReturn  (n=35)
    Pearson  r=-0.0092  p=0.9581
    Spearman r=+0.0726  p=0.6785

  Abl 10, treatment (n=35):
  rho_bar vs Sharpe  (n=35)
    Pearson  r=+0.1059  p=0.5449
    Spearman r=+0.0540  p=0.7582

  rho_bar vs TotalReturn  (n=35)
    Pearson  r=+0.0858  p=0.6243
    Spearman r=+0.0142  p=0.9357


========================================================================
4. WITHIN SCENARIO PAIRED DIFFERENCES (delta analysis)
========================================================================
For each scenario appearing in both baseline and treatment within an ablation,
compute delta_rho = treatment.rho_bar - baseline.rho_bar
         delta_sharpe = treatment.sharpe - baseline.sharpe
         delta_return = treatment.total_return - baseline.total_return
Then correlate the deltas.

Paired scenarios found: 105
  Ablation 7: 35 pairs
  Ablation 8: 35 pairs
  Ablation 10: 35 pairs


--- Pooled deltas ---
  delta_rho:    mean=-0.0003  std=0.0329
  delta_sharpe: mean=+0.0327  std=0.4072
  delta_return: mean=+0.2622  std=2.0966

  delta_rho vs delta_Sharpe  (n=105)
    Pearson  r=-0.0157  p=0.8738
    Spearman r=+0.0066  p=0.9465

  delta_rho vs delta_Return  (n=105)
    Pearson  r=+0.0114  p=0.9084
    Spearman r=+0.0331  p=0.7373


--- Deltas within ablation 7 ---
  n=35 pairs
  delta_rho:    mean=+0.0013  std=0.0382
  delta_sharpe: mean=+0.1432  std=0.3684
  delta_return: mean=+0.8285  std=2.4367

  delta_rho vs delta_Sharpe  (n=35)
    Pearson  r=-0.1270  p=0.4673
    Spearman r=-0.1219  p=0.4856

  delta_rho vs delta_Return  (n=35)
    Pearson  r=-0.0690  p=0.6935
    Spearman r=-0.0521  p=0.7663


--- Deltas within ablation 8 ---
  n=35 pairs
  delta_rho:    mean=-0.0139  std=0.0234
  delta_sharpe: mean=+0.0439  std=0.4085
  delta_return: mean=+0.1557  std=1.8887

  delta_rho vs delta_Sharpe  (n=35)
    Pearson  r=-0.0109  p=0.9505
    Spearman r=+0.0132  p=0.9402

  delta_rho vs delta_Return  (n=35)
    Pearson  r=+0.0720  p=0.6809
    Spearman r=+0.0565  p=0.7474


--- Deltas within ablation 10 ---
  n=35 pairs
  delta_rho:    mean=+0.0116  std=0.0312
  delta_sharpe: mean=-0.0890  std=0.4203
  delta_return: mean=-0.1975  std=1.8389

  delta_rho vs delta_Sharpe  (n=35)
    Pearson  r=+0.2072  p=0.2324
    Spearman r=+0.2217  p=0.2005

  delta_rho vs delta_Return  (n=35)
    Pearson  r=+0.1545  p=0.3755
    Spearman r=+0.1266  p=0.4685


========================================================================
5. PER AGENT RHO CORRELATIONS
========================================================================
Agents found across dataset: macro, risk, technical


--- Agent: macro ---
  n=210 runs with rho_macro
  rho_macro vs Sharpe  (n=210)
    Pearson  r=+0.0377  p=0.5866
    Spearman r=+0.0212  p=0.7602

  rho_macro vs TotalReturn  (n=210)
    Pearson  r=+0.0158  p=0.8202
    Spearman r=+0.0017  p=0.9810


--- Agent: risk ---
  n=70 runs with rho_risk
  rho_risk vs Sharpe  (n=70)
    Pearson  r=+0.0083  p=0.9453
    Spearman r=+0.0027  p=0.9826

  rho_risk vs TotalReturn  (n=70)
    Pearson  r=+0.0405  p=0.7393
    Spearman r=+0.0084  p=0.9448


--- Agent: technical ---
  n=210 runs with rho_technical
  rho_technical vs Sharpe  (n=210)
    Pearson  r=+0.0820  p=0.2366
    Spearman r=+0.0706  p=0.3086

  rho_technical vs TotalReturn  (n=210)
    Pearson  r=+0.0274  p=0.6932
    Spearman r=+0.0464  p=0.5033


--- Agent rho comparison table (Pearson r with Sharpe) ---
  Agent                  r        p     n
  --------------------------------------
  macro            +0.0377   0.5866   210
  risk             +0.0083   0.9453    70
  technical        +0.0820   0.2366   210


--- Agent rho comparison table (Pearson r with TotalReturn) ---
  Agent                  r        p     n
  --------------------------------------
  macro            +0.0158   0.8202   210
  risk             +0.0405   0.7393    70
  technical        +0.0274   0.6932   210

========================================================================
6. OLS REGRESSION: Sharpe ~ rho_bar + ablation dummies
========================================================================
Controls for ablation level effects using dummy variables.
Reference ablation: the lowest numbered ablation.


--- Dependent variable: Sharpe (ref ablation=7) ---
  n=210, R2=0.0072, Adj R2=-0.0072
  F(3,206)=0.499, p=0.6832
  Variable                        Coef         SE        t        p
  ---------------------------------------------------------------
  (intercept)                  -3.8738     4.2549   -0.910   0.3637
  rho_bar                       5.3388     5.1030    1.046   0.2967
  abl_8_dummy                  -0.1515     0.2803   -0.540   0.5895
  abl_10_dummy                 -0.0064     0.2829   -0.023   0.9820


--- Dependent variable: TotalReturn (ref ablation=7) ---
  n=210, R2=0.0028, Adj R2=-0.0117
  F(3,206)=0.196, p=0.8993
  Variable                        Coef         SE        t        p
  ---------------------------------------------------------------
  (intercept)                  -5.4973    22.1532   -0.248   0.8043
  rho_bar                       9.2436    26.5692    0.348   0.7283
  abl_8_dummy                  -0.7028     1.4593   -0.482   0.6306
  abl_10_dummy                  0.2261     1.4732    0.153   0.8782


--- Extended model: Sharpe ~ rho_bar + ablation dummies + condition dummy ---
  n=210, R2=0.0073, Adj R2=-0.0120
  F(4,205)=0.378, p=0.8239
  Variable                        Coef         SE        t        p
  ---------------------------------------------------------------
  (intercept)                  -3.8961     4.2675   -0.913   0.3623
  rho_bar                       5.3447     5.1153    1.045   0.2973
  abl_8_dummy                  -0.1515     0.2810   -0.539   0.5904
  abl_10_dummy                 -0.0064     0.2836   -0.023   0.9819
  treatment_dummy               0.0345     0.2288    0.151   0.8801


--- Extended model: TotalReturn ~ rho_bar + ablation dummies + condition dummy ---
  n=210, R2=0.0031, Adj R2=-0.0164
  F(4,205)=0.158, p=0.9590
  Variable                        Coef         SE        t        p
  ---------------------------------------------------------------
  (intercept)                  -5.6681    22.2177   -0.255   0.7989
  rho_bar                       9.2893    26.6315    0.349   0.7276
  abl_8_dummy                  -0.7030     1.4627   -0.481   0.6313
  abl_10_dummy                  0.2257     1.4766    0.153   0.8787
  treatment_dummy               0.2654     1.1912    0.223   0.8239


========================================================================
7. SUMMARY AND INTERPRETATION
========================================================================
Total observations: 210
Pooled rho_bar Sharpe:      Pearson r=+0.0735 (p=0.2887), Spearman r=+0.0481 (p=0.4879)
Pooled rho_bar TotalReturn: Pearson r=+0.0265 (p=0.7030)

No statistically significant pooled correlation between CRIT scores and Sharpe.

Delta analysis (paired within scenario): r=-0.0157 (p=0.8738)
Changes in CRIT quality do NOT significantly predict changes in financial performance.

Note: Correlations marked with * are significant at p < 0.05.
All p values are two tailed.
\end{lstlisting}

\subsection{Causal Debate Baselines \& RCA}
\label{app:debate_rca}

\begin{table}[h]
\centering\small
\begin{tabular}{llll}
\toprule
Metric & \shortstack{(A) 3 Roles \\ Mean, Causal} & \shortstack{(B) 3 Roles \\ Debate, Causal} & \shortstack{(C) 3 Roles \\ Debate, Causal + RCA} \\
\midrule
Excess Return \% (v SPY) & $1.12\% \pm 0.93\%$ & $0.95\% \pm 0.96\%$ & $0.80\% \pm 0.88\%$ \\
 Sharpe & $1.07 \pm 0.30$ & $1.01 \pm 0.30$ & $1.03 \pm 0.30$ \\
 Volatility & ${0.21 \pm 0.01}^{\mathbf{B}\mathbf{C}}$ & ${0.20 \pm 0.01}^{\mathbf{A}C}$ & ${0.19 \pm 0.01}^{\mathbf{A}B}$ \\
Max Drawdown \% & ${9.33\% \pm 0.90\%}^{B\mathbf{C}}$ & ${8.88\% \pm 0.83\%}^{AC}$ & ${8.64\% \pm 0.80\%}^{\mathbf{A}B}$ \\
 Sortino & $1.87 \pm 0.47$ & $1.82 \pm 0.49$ & $1.80 \pm 0.47$ \\
% Calmar Ratio & $6.09 \pm 1.40$ & $6.09 \pm 1.50$ & $5.98 \pm 1.43$ \\
Num Positions & $10.83 \pm 1.41$ & $11.17 \pm 1.49$ & $11.14 \pm 1.57$ \\
Uninvested Cash & ${7.51\% \pm 0.57\%}^{\mathbf{B}\mathbf{C}}$ & ${12.63\% \pm 0.78\%}^{\mathbf{A}C}$ & ${13.80\% \pm 1.00\%}^{\mathbf{A}B}$ \\
Total Return \% & $4.63\% \pm 1.54\%$ & $4.47\% \pm 1.49\%$ & $4.31\% \pm 1.41\%$ \\
\bottomrule
\end{tabular}
\vspace{4pt}
\caption{Mean over 35 quarter+tickers scenarios $\pm$ SEM. Superscripts indicate stat.\ sig.\ in paired t-test with the specified column(s). Unbolded: $p < 0.15$, \textbf{Bolded}: $p < 0.01$).}
\label{tab:debate_rca}
\end{table}

\subsection{Enriched Debate Scenario Breakdown}
\label{app:per-scenario-debate}

\begin{table}[h]
\centering
\small
\begin{tabular}{@{}lcccccc@{}}
\toprule
\textbf{Scenario} & \textbf{Return \%} & \textbf{Sharpe} & \textbf{Sortino} & \textbf{Max DD \%} & \textbf{SPY \%} & \textbf{Excess \%} \\
\midrule
2022\,Q4 Tech Sector      & $-0.35$ & $-0.13$ & $-0.20$ & $5.96$ & $+4.79$ & $-5.14$ \\
2022\,Q1 Inflation Shock  & $+2.83$ & $0.89$  & $1.32$  & $6.11$ & $-5.16$ & $+7.99$ \\
2023\,Q2 Higher-for-Longer & $+5.52$ & $1.52$  & $2.37$  & $3.91$ & $+8.27$ & $-2.75$ \\
\midrule
\textbf{Average}          & $+2.67$ & $0.76$  & $1.16$  & $5.33$ & $+2.63$ & $+0.04$ \\
\bottomrule
\end{tabular}
\vspace{4pt}
\setlength{\belowcaptionskip}{-4pt}
\caption{Per-scenario results for 3-agent enriched debate (Config~3). Performance varies substantially across market regimes: the system outperforms SPY by $+7.99\%$ during the 2022\,Q1 inflation shock but underperforms by $-5.14\%$ in the 2022\,Q4 tech rally.}
\label{tab:per-scenario-debate}
\end{table}

\subsection{Reasoning Quality}
\label{app:crit-scores}

\begin{table}[H]
\centering
\small
\begin{tabular}{@{}lccccc@{}}
\toprule
\textbf{Round} & $\bar{\rho}$ & \textbf{$P_1$ (LV)} & \textbf{$P_2$ (ES)} & \textbf{$P_3$ (AC)} & \textbf{$P_4$ (CA)} \\
\midrule
1 & 0.843 & 0.86 & 0.86 & 0.87 & 0.78 \\
2 & 0.845 & 0.87 & 0.85 & 0.86 & 0.81 \\
3 & 0.815 & 0.84 & 0.83 & 0.80 & 0.75 \\
4 & 0.832 & 0.77 & 0.86 & 0.88 & 0.81 \\
5 & 0.839 & 0.82 & 0.86 & 0.86 & 0.79 \\
\bottomrule
\end{tabular}
\vspace{4pt}
\setlength{\belowcaptionskip}{-4pt}
\caption{CRIT pillar scores across debate rounds (4-agent enriched debate, 2022\,Q4 tech scenario). $P_1$=Logical Validity, $P_2$=Evidential Support, $P_3$=Alternative Consideration, $P_4$=Causal Alignment. The most common CRIT diagnostic is \texttt{causal\_overreach}, particularly for the Technical agent.}
\label{tab:crit-scores}
\end{table}

\subsection{GPT and Gemini Experiments}
\label{app:llm-comparison}

\begin{table}[h]
\centering
\small
\begin{tabular}{@{}llcccccc@{}}
\toprule
\textbf{Scenario} & \textbf{Exp} & \textbf{Return \%} & \textbf{Sharpe} & \textbf{Sortino} & \textbf{Max DD \%} & \textbf{Excess \%} & \textbf{Status} \\
\midrule
2021\,Q3 Inflation   & A & $+12.40$ & $3.05$ & $4.65$ & $4.61$ & $+2.64$ & OK \\
                      & B & -       & -     & -     & -     & -      & \textsc{Fail} \\
\midrule
2022\,Q1 Inf.\ Shock & A & $-19.40$ & $-2.56$ & $-3.16$ & $19.96$ & $-3.05$ & OK \\
                      & B & $-19.40$ & $-2.56$ & $-3.16$ & $19.96$ & $-3.05$ & OK \\
\midrule
2022\,Q2 Recession   & A & $-1.97$  & $-0.35$ & $-0.50$ & $12.69$ & $+3.96$ & OK \\
                      & B & $-1.77$  & $-0.32$ & $-0.45$ & $12.97$ & $+4.16$ & OK \\
\midrule
2022\,Q4 Banking     & A & $+2.75$  & $0.49$  & $0.70$  & $8.16$  & $-5.17$ & OK \\
                      & B & -       & -      & -      & -      & -      & \textsc{Fail} \\
\midrule
2023\,Q2 Higher4Lngr & A & $-0.03$  & $-0.37$ & $-0.50$ & $5.53$  & $+3.31$ & OK \\
                      & B & -       & -      & -      & -      & -      & \textsc{Fail} \\
\midrule
2023\,Q4 AI Boom     & A & $+16.91$ & $4.66$  & $9.24$  & $2.04$  & $+5.90$ & OK \\
                      & B & $+14.42$ & $4.43$  & $8.40$  & $1.91$  & $+3.41$ & OK \\
\midrule
\textbf{Mean (completed)} & A & $+1.78$ & $0.82$ & $1.74$ & $8.83$ & $+1.26$ & 6/6 \\
                           & B & $-2.25$ & $0.52$ & $1.60$ & $11.61$ & $+1.51$ & 3/6 \\
\bottomrule
\end{tabular}
\vspace{4pt}
\setlength{\belowcaptionskip}{-4pt}
\caption{Per-scenario LLM comparison. Exp~A uses GPT-5-mini for all roles; Exp~B uses Gemini-2.5-flash for Macro/Value/Judge and GPT-5-mini for Risk. Exp~B failures are caused by Gemini hallucinating ticker JNJ (not in scenario universe). Where both complete, returns are comparable- e.g., 2022\,Q1 produces byte-identical portfolios.}
\label{tab:llm-comparison}
\end{table}

\subsection{GPT and Gemini Phase LLM results}
\begin{table}[h]
\centering
\small
\begin{tabular}{@{}lc@{}}
\toprule
\textbf{Metric} & \textbf{Value} \\
\midrule
Scenario         & 2022\,Q1 Inflation Shock (10 tickers) \\
Return           & $-16.22\%$ \\
Sharpe           & $-2.11$ \\
Sortino          & $-2.63$ \\
Max Drawdown     & $17.67\%$ \\
SPY Return       & $-16.35\%$ \\
Excess vs SPY    & $+0.13\%$ \\
\bottomrule
\end{tabular}
\vspace{4pt}
\setlength{\belowcaptionskip}{-4pt}
\caption{Phase-LLM experiment results (GPT-5-mini for propose/revise, Gemini-2.5-flash critique/judge). Tracked SPY closely during a severe downturn, with excess return of $+0.13\%$.}
\label{tab:phase-llm}
\end{table}

\subsection{GPT and Anthropic Ablation Tests}
\label{app:ablation_performance}

\begin{table}[H]
\centering
\begin{tabular}{lccc}
\hline
\textbf{Metric} & \textbf{Mean} & \textbf{Best} & \textbf{Worst} \\
\hline
Return (\%) & -1.11 & 21.71 & -70.10 \\
Excess Return vs SPY (\%) & 1.01 & 11.58 & -6.41 \\
Max Drawdown (\%) & 7.87 & 1.68 & 16.39 \\
Sharpe Ratio & 1.06 & 4.38 & -2.73 \\
Trades & 8.8 & 21 & 1 \\
\hline
\end{tabular}
\vspace{4pt}
\caption{Ablation performance across 17 market scenarios.}
\label{tab:ablation_performance}
\end{table}

\subsection{GPT and Anthropic Routing Experiments}
\label{app:llm_routing_experiments}

\begin{table}[H]
\centering
\begin{tabular}{lcccc}
\hline
\textbf{Experiment} & \textbf{Return (\%)} & \textbf{Excess vs SPY (\%)} & \textbf{Sharpe} & \textbf{Final JS} \\
\hline
Exp 1 Role-only (Macro Anthropic, Risk GPT) & -1.42 & 3.16 & -0.54 & 0.7181 \\
Exp 2 Role-only (Macro GPT, Risk Anthropic) & -2.37 & 2.26 & -0.64 & 0.7312 \\
Exp 3 Hybrid phases & -1.70 & 2.84 & -0.53 & 0.7946 \\
\hline
\end{tabular}
\vspace{4pt}
\caption{Performance comparison across LLM routing experiments.}
\label{tab:llm_routing_experiments}
\end{table}

\subsection{Performance Metrics across Intervention Ablations}
\label{app:financial-results}

\begin{table}[H]
\centering
\small
\begin{tabular}{llrrrrrr}
\toprule
\textbf{Abl.} & \textbf{Config} & \textbf{N} & \textbf{Ret} & \textbf{Sharpe} & \textbf{Sortino} & \textbf{MaxDD} & \textbf{Excess vs SPY} \\
\midrule
\multirow{2}{*}{A7}
& Base & 35 & $+1.79 \pm 1.50$ & $+0.50 \pm 0.28$ & $+0.93 \pm 0.43$ & $+10.03 \pm 0.88$ & $-1.73 \pm 1.03$ \\
& +JS & 35 & $+2.62 \pm 1.61$ & {\color{darkgreen}$\mathbf{+0.64 \pm 0.29}$} & {\color{darkgreen}$\mathbf{+1.19 \pm 0.46}$} & $+10.21 \pm 0.95$ & {\color{darkgreen}$-0.90 \pm 1.08$} \\
\midrule
\multirow{2}{*}{A8}
& Base & 35 & $+1.46 \pm 1.21$ & $+0.42 \pm 0.28$ & $+0.81 \pm 0.44$ & $+8.07 \pm 0.71$ & $-2.06 \pm 0.87$ \\
& +CRIT & 35 & $+1.61 \pm 1.26$ & $+0.46 \pm 0.28$ & $+0.86 \pm 0.43$ & $+8.27 \pm 0.78$ & $-1.90 \pm 0.94$ \\
\midrule
\multirow{2}{*}{A10}
& Base & 35 & $+2.61 \pm 1.54$ & $+0.66 \pm 0.28$ & $+1.18 \pm 0.44$ & $+9.80 \pm 0.87$ & $-0.91 \pm 1.04$ \\
& +JS & 35 & $+2.41 \pm 1.59$ & $+0.57 \pm 0.27$ & $+1.04 \pm 0.42$ & {\color{red}$\mathbf{+10.33 \pm 0.93}$} & $-1.11 \pm 1.07$ \\
\bottomrule
\end{tabular}
\vspace{6pt}
\caption{Performance metrics across intervention ablations. Three ablations evaluate reasoning interventions in the multi-agent debate system. \textbf{A7} introduces a Jensen--Shannon (JS) collapse detector in a two-agent setting 
(\textit{macro}, \textit{technical}), triggering when revision-stage disagreement falls below a threshold 
($JS_{\text{rev}} / JS_{\text{prop}} < 0.8$) and nudging agents to maintain independent reasoning. 
\textbf{A8} applies the same JS intervention with an added \textit{risk} agent (three agents total); 
the intervention fired frequently (97\% of debates) but produced no measurable effect. 
\textbf{A10} replaces the JS trigger with a CRIT-based reasoning intervention that fires when 
causal reasoning quality $(\rho_4 < 0.8)$.}
\vspace{-10pt}
\label{tab:financial-results}

\end{table}

\subsection{T-tests Results for Ablations}
\label{app:paired-ttests}

\begin{table}[H]
\centering
\small
\begin{tabular}{@{}llccccc@{}}
\toprule
\textbf{Abl.} & \textbf{Metric} & \textbf{Diff} & \textbf{95\% CI} & \textbf{t} & \textbf{p} & \textbf{d} \\
\midrule
A7 & Sharpe           & $+0.143$ & $[+0.017,\; +0.270]$ & $+2.300$ & \textbf{\textcolor{darkgreen}{0.0277}} & $+0.389$ \\
A7 & Sortino          & $+0.254$ & $[+0.032,\; +0.476]$ & $+2.328$ & \textbf{\textcolor{darkgreen}{0.0260}} & $+0.393$ \\
A7 & Excess vs SPY \% & $+0.828$ & $[-0.009,\; +1.666]$ & $+2.011$ & \textcolor{darkgreen}{0.0523} & $+0.340$ \\
A10 & Max DD \%       & $+0.529$ & $[-0.028,\; +1.086]$ & $+1.930$ & \textbf{\textcolor{red}{0.0620}} & $+0.326$ \\
\bottomrule
\end{tabular}
\vspace{6pt}
\caption{Paired t-test results for ablations with statistically significant or marginal effects ($p < 0.10$). Green p-values indicate improvement relative to the baseline configuration, red indicates undesired outcomes.}
\label{tab:paired-ttests}
\end{table}

\subsection{Prompt Enrichment Tiers}
\label{app:prompt_tiers}

\begin{table}[H]
\centering
\small
\begin{tabular}{lcccc}
\toprule
\textbf{Tier} & \textbf{n} & $\boldsymbol{\bar{\rho}}$ & \textbf{Sharpe} & \textbf{Return (\%)} \\
\midrule
Standard        & 56 & 0.833 & -0.918 & -3.56 \\
Light           & 62 & 0.825 & -0.940 & -3.58 \\
Intense Light   & 67 & 0.819 & -0.992 & -3.66 \\
Intense         & 60 & 0.834 & -0.967 & -3.75 \\
\bottomrule
\end{tabular}
\vspace{4pt}
\caption{Prompt enrichment tiers (Standard $\rightarrow$ Intense) show no improvement in reasoning or financial performance across 245 runs of Ablation~1.}
\label{tab:prompt_tiers}
\end{table}

\subsection{Pearson correlation between Agent Reasoning Quality and Performance}
\label{app:correlation_betwen_agents}

\begin{table}[H]
\centering
\small
\begin{tabular}{lcccc}
\toprule
\textbf{Subgroup} & \textbf{Metric} & \textbf{r} & \textbf{p} & \textbf{n} \\
\midrule
Value agent $\rho$ vs Sharpe & Pearson & $-0.29$ & 0.001 & 126 \\
Risk agent $\rho$ vs Return & Pearson & $+0.22$ & 0.006 & 152 \\
\bottomrule
\end{tabular}
\vspace{4pt}
\caption{Selected subgroup correlations between agent reasoning quality and performance.}
\end{table}

\subsection{Debate Portfolio Performance}
\label{app:debate_portofolio}
\begin{table}[H]
\centering
\small
\begin{tabular}{lccccc}
\toprule
\textbf{Ablation} & \textbf{n} & \textbf{Debate Return} & \textbf{EW Return} & \textbf{Diff} & \textbf{Beats EW} \\
\midrule
Abl 1 & 245 & $-3.64\%$ & $-3.28\%$ & $-0.36\%$ & 31\% \\
Abl 7 & 70 & $+2.20\%$ & $+4.47\%$ & $-2.26\%$ & 30\% \\
Abl 8 & 70 & $+1.54\%$ & $+4.47\%$ & $-2.93\%$ & 34\% \\
Abl 10 & 70 & $+2.51\%$ & $+4.47\%$ & $-1.96\%$ & 40\% \\
\textbf{Pooled} & 455 & --- & --- & $\mathbf{-1.29\%}$ & \textbf{33\%} \\
\bottomrule
\end{tabular}
\vspace{4pt}
\caption{Debate portfolio performance compared to an equal-weight baseline.}
\end{table}

\subsection{Improvements in CRIT}
\label{app:improvement_in_crit}
\begin{table}[H]
\centering
\small
\begin{tabular}{lcccc}
\toprule
\textbf{Pillar} & \textbf{Baseline} & \textbf{Enriched} & $\Delta$ & \textbf{Cohen's $d$} \\
\midrule
Evidential Support & 0.51 & 0.85 & $+0.34$ & 2.92 \\
Causal Alignment & 0.70 & 0.87 & $+0.17$ & 2.21 \\
Alternative Consideration & 0.79 & 0.89 & $+0.10$ & 1.09 \\
Logical Validity & 0.87 & 0.82 & $-0.05$ & $-0.44$ \\
\bottomrule
\end{tabular}
\vspace{4pt}
\caption{Improvements in CRIT reasoning pillars under structured prompts.}
\end{table}

% \newpage
% \section{Appendix - Prompts}

% \appendix
% \section{Two Key Prompts: CRIT, Enriched Agent Roles}
\subsection{CRIT Prompt}
\label{app:crit-prompt}

\begin{lstlisting}[backgroundcolor=\color{promptbg}]

## Round {{ round }}

## Agent Under Evaluation

{{ agent_role | upper }}

Your task is to evaluate the reasoning integrity of a trading agent's REVISED ARGUMENT after a debate round.

CRIT evaluates reasoning quality only.

You must NOT evaluate:

- prediction accuracy
- trading profitability
- market outcomes
- agreement with other agents

You do not know the correct investment decision.

Your evaluation must focus strictly on the soundness of the reasoning process.

---

# Evaluation Target

You are provided:

- the original proposal
- critiques from other agents
- the revised argument

Use the proposal and critiques only to determine:

- whether critiques exposed reasoning weaknesses
- whether the revision addressed those weaknesses
- whether competing explanations were considered

The primary object of evaluation is the revised argument.

---

# Agent Specialization

Agent specialization: {{ agent_role }}

This agent focuses on {{ agent_role }} reasoning.

Evaluate reasoning within this specialization.

Do NOT penalize arguments for focusing on their domain expertise.

---

# Structured Reasoning Format

The revised argument contains structured reasoning fields.

reasoning.claims

Each claim includes:

- claim_id
- claim_text
- claim_type (macro / sector / firm / risk / technical)
- reasoning_type (causal / observational / risk_assessment / pattern)
- evidence_ids (normalized evidence references)
- assumptions
- falsifiers
- impacts_positions (tickers affected by this claim)
- confidence

reasoning.position_rationale

Per-position explanations containing:

- ticker
- weight
- supporting_claims (claim IDs that justify the position)
- explanation

reasoning.risks_or_falsifiers

Portfolio-level invalidation conditions.

reasoning.revision_notes

Explanation of how critiques were handled.

reasoning.critique_responses

Per-critique structured responses. Each entry contains:

- from_agent (which agent sent the critique)
- target_claim (which claim was critiqued)
- disposition ("accept" or "rebut")
- justification (why the agent accepted or rebutted)

Use this to determine whether each critique was explicitly addressed.

A critique is IGNORED only if no matching critique_response entry exists for it.

portfolio_allocation

Final asset allocation.

evidence_citations

Full text for evidence references.

Important distinctions:

- claims[*].impacts_positions -> which assets the claim theoretically affects
- position_rationale[*].supporting_claims -> which claims justify the allocation

These serve different purposes:

- impacts_positions -> causal completeness
- supporting_claims -> conclusion consistency

Important note about impacts_positions:

The impacts_positions field may legitimately be empty.

Some claims describe macro or market-wide relationships that affect the portfolio indirectly
(e.g., interest rates affecting equity valuations broadly).

Absence of impacts_positions is NOT automatically a reasoning failure.

Only penalize causal completeness when a specific portfolio position is clearly implied
by a claim but the reasoning fails to connect that claim to the position.

---

# Debate Structure

Arguments are structured using explicit claims.

Claims are labeled:

C1, C2, C3, ...

Critiques may target specific claims or the portfolio allocation.

During revision agents may:

- ACCEPT critiques
- REJECT critiques with justification

Your task is to evaluate whether the revised reasoning properly engages with critiques.

---

# Original Proposal

Use the proposal only to determine whether critiques revealed weaknesses.

{{ proposal | tojson(indent=2) }}

---

# Critiques Received

Each critique contains:

- from_role - which agent sent the critique
- target_claim - which specific claim (C1, C2, ...) is targeted
- critique_text - the objection
- evidence_citations - counter-evidence IDs with full text
- portfolio_implication - how the flaw affects the portfolio
- suggested_adjustment - what the critic recommends
- falsifier - what would disprove the objection
- objection_confidence - how confident the critic is

Use target_claim to match critiques against critique_responses in the revised argument.

{% if critiques_received | length == 0 %}
No critiques were received.

When critiques are absent:

- Do NOT penalize alternative_consideration for critique handling
- Evaluate whether the agent independently considers alternative explanations
{% else %}
Evaluate whether critiques were meaningfully addressed in the revision.
{% endif %}

{{ critiques_received | tojson(indent=2) }}

---

{% if self_critique is defined and self_critique is not none %}
# Agent Self-Critique

The agent identified the following weakness in its own reasoning during the critique phase:

{{ self_critique | tojson(indent=2) }}

Evaluate whether the revision addresses the agent's own identified weakness.

If the agent identified a weakness but the revision ignores it, this is relevant
to alternative_consideration.

Do not reward agents merely for writing a self-critique.

Only reward reasoning improvements that appear in the revised argument.

---

{% endif %}

# Revised Argument

Focus primarily on the reasoning quality in this revision.

{{ revised_argument | tojson(indent=2) }}

---

# Evaluation Method

Your evaluation must follow two phases.

PHASE 1 - Detection  
Identify objective reasoning failures.

PHASE 2 - Judgment  
Assign pillar scores based on detected failures.

Detection should be conservative and objective.

Scores must reflect the number and severity of detected failures.

Scores MUST be consistent with the failure counts detected in Phase 1.

---

# PHASE 1 - Detect Reasoning Failures

Before assigning scores, detect the following reasoning failures.

Record counts where possible.

---

## Logical Failures

Detect:

- contradictions within the argument
- mutually incompatible claims
- circular reasoning
- missing reasoning steps between evidence and conclusions
- conclusions that do not logically follow from earlier reasoning

Also detect claim-to-claim contradictions, where two claims imply incompatible macro or market outcomes.

Example:

- one claim implies recession
- another implies strong cyclical expansion

Also evaluate whether the portfolio allocation logically follows from the thesis.

If contradictions exist, count them.

---

## Evidence Failures

Evaluate how evidence supports claims.

Cross-reference:

reasoning.claims[*].evidence_ids with evidence_citations.

If evidence_ids is present and non-empty on a claim, use it as the canonical evidence reference.

If evidence_ids is empty or absent, check whether the claim's claim_text contains
inline bracketed references (e.g., [AAPL-F2], [L1-FF]).

If inline references exist but evidence_ids is empty, treat the claim as having evidence.

A claim is unsupported only when it has no evidence_ids AND no inline bracketed references in claim_text.

Let:

total_claims = number of claims in reasoning.claims  
unsupported_claims = number of claims failing the evidence rule

unsupported_claim_rate = unsupported_claims / total_claims

Detect:

- claims with no evidence citations
- citations that do not support the claim
- misinterpretation of evidence
- claims stronger than the cited evidence justifies

Count unsupported claims.

---

## Alternative Reasoning Failures

Evaluate how the agent handles competing explanations and critiques.

Detect:

- ignored critiques
- critiques rejected without justification
- accepted critiques that produce no meaningful reasoning change
- failure to consider alternative explanations
- premature certainty

A critique is considered ADDRESSED only if the revision provides
explicit reasoning responding to the critique.

Primary source: reasoning.critique_responses.

Cross-reference these against critiques_received to identify ignored critiques.

Secondary evidence of addressing may include:

- revision_notes explaining why a critique is accepted or rejected
- claim structure changes motivated by the critique
- portfolio allocation changes motivated by the critique
- new evidence introduced to evaluate the critique

Superficial acknowledgement without reasoning does NOT count.

If a critique has no matching entry in critique_responses AND no secondary evidence
of engagement, treat the critique as ignored.

---

## Causal Reasoning Failures

Evaluate causal reasoning using Pearl's causal hierarchy.

For each claim determine its causal level:

L1 - correlation / association  
L2 - intervention / mechanism  
L3 - counterfactual reasoning

Detect rung collapse including:

- L2 claims without mechanism description
- L3 claims without counterfactual reasoning
- conclusions requiring L2/L3 reasoning but supported only by L1 evidence
- macro claims mapped to asset outcomes without transmission mechanisms

Count causal overreach cases.

---

## Assumption Failures

Detect:

- critical assumptions not acknowledged
- assumptions contradicted by evidence
- assumptions doing most explanatory work

These typically reduce evidential_support.

---

## Conclusion Consistency Failures

Determine whether the portfolio allocation follows logically from the reasoning.

Use:

- position_rationale[*].supporting_claims
- claims[*].impacts_positions

Detect:

- positions >10% with no supporting_claims
- supporting_claims referencing missing claim IDs
- claims that do not logically imply the allocation
- portfolio changes without reasoning explanation

Count orphaned positions.

---

# Failure -> Pillar Mapping

logical_validity penalties:

- contradictions
- missing reasoning steps
- portfolio reasoning mismatch

evidential_support penalties:

- unsupported claims
- evidence misuse

alternative_consideration penalties:

- ignored critiques
- unjustified rejection of critiques
- premature certainty

causal_alignment penalties:

- causal overreach
- missing causal mechanisms

---

# PHASE 2 - Assign Pillar Scores

Assign scores for the four CRIT pillars.

Scores must be consistent with detected counts.

---

# Score Interpretation

Scores must be within 0.0 - 1.0.

1.00 - excellent reasoning  
0.85 - strong reasoning with minor weaknesses  
0.70 - acceptable reasoning with noticeable gaps  
0.55 - mixed reasoning with clear problems  
0.40 - weak reasoning with major flaws  
0.20 - very weak reasoning  
0.00 - severe reasoning failure

---

# Score Constraints

Apply these rules when failures are detected:

- contradictions -> logical_validity <= 0.70
- unsupported claims -> evidential_support <= 0.70
- ignored critiques -> alternative_consideration <= 0.70
- causal overreach -> causal_alignment <= 0.70

Severity guidance:

- 1 failure -> small penalty
- 2-3 failures -> moderate penalty (typically <= 0.55)
- 4+ failures -> severe penalty (typically <= 0.40)

Scores above 0.90 should be rare and require near-perfect reasoning.

Avoid defaulting to simple values such as 0.5 or 1.0 unless fully justified.

---

# Diagnostic Boolean Rules

unsupported_claims_detected = unsupported_claims_count > 0  
ignored_critiques_detected = ignored_critiques_count > 0  
causal_overreach_detected = causal_overreach_count > 0  
contradictions_detected = contradictions_count > 0  

---

# Output

Return only the JSON object below.

No markdown.  
No text outside JSON.

{
  "pillar_scores": {
    "logical_validity": <float>,
    "evidential_support": <float>,
    "alternative_consideration": <float>,
    "causal_alignment": <float>
  },
  "diagnostics": {
    "contradictions_detected": <bool>,
    "unsupported_claims_detected": <bool>,
    "ignored_critiques_detected": <bool>,
    "premature_certainty_detected": <bool>,
    "causal_overreach_detected": <bool>,
    "conclusion_drift_detected": <bool>,
    "contradictions_count": <int or null>,
    "unsupported_claims_count": <int or null>,
    "ignored_critiques_count": <int or null>,
    "causal_overreach_count": <int or null>,
    "orphaned_positions_count": <int or null>
  },
  "explanations": {
    "logical_validity": "<1-2 sentences explaining key reasoning issues>",
    "evidential_support": "<1-2 sentences explaining evidence usage>",
    "alternative_consideration": "<1-2 sentences explaining critique handling>",
    "causal_alignment": "<1-2 sentences explaining causal reasoning quality>"
  }
}
\end{lstlisting}

\subsection{Macro Role Basic Prompt}
\label{app:macro-basic}
\begin{lstlisting}
[backgroundcolor=\color{promptbg}]

You are the MACRO STRATEGIST agent on a multi-agent trading desk.

## Your Analytical Domain
You specialize in macroeconomic analysis and its impact on equity prices:
- Federal Reserve policy: interest rate decisions, forward guidance, QE/QT
- Inflation indicators: CPI, PCE, PPI, inflation expectations
- Economic growth: GDP, employment data, PMI, consumer confidence
- Yield curve dynamics: term spreads, real rates, breakeven inflation
- Global macro: currency movements, trade flows, geopolitical risk
- Fiscal policy: government spending, tax policy, deficit trajectory

## How to Reason (Step by Step)
1. Identify the dominant macro regime (expansion, contraction, transition)
2. Map macro signals to equity sector and factor implications
3. Explicitly state what macro data release would falsify your view
4. Consider: could the market have already priced in this macro signal?

## Evidence Citation - MANDATORY
You MUST cite data using ONLY the exact bracketed evidence IDs from the memo.
Valid citations look like: [L1-FF], [L1-VIX], [L1-10Y], [AAPL-BETA].
NEVER invent your own evidence labels. NEVER paraphrase IDs as free text like
"fedrates" or "inflationdata" - these are INVALID and will be rejected.
If you reference a data point, find its exact [TICKER-METRIC] or [L1-METRIC] tag
in the memo and use that tag verbatim. Any claim without a valid bracketed evidence
ID is unsupported.

## Output Format
Respond with valid JSON only (no markdown, no extra text).

CRITICAL CONSTRAINT - ALLOWED TICKERS:
You may ONLY allocate to the tickers listed in the "Allocation universe" above.
Do NOT include any other tickers. Any ticker not in that list will be discarded.
Your JSON "allocation" object must contain ONLY tickers from the allocation universe,
and it must contain ALL of them (use 0.0 for tickers you do not want to hold).

Example - if the universe is [AAPL, NVDA, MSFT]:
{
  "justification": "...",
  "risks_or_falsifiers": "...",
  "allocation": {"AAPL": 0.40, "NVDA": 0.35, "MSFT": 0.25},
  "confidence": 0.75
}

WRONG (ticker not in universe):
{
  "allocation": {"AAPL": 0.30, "NFLX": 0.30, "MSFT": 0.40}
}
NFLX is not in the universe - this is INVALID.

Full JSON schema:
{
  "justification": "Your reasoning for this allocation",
  "risks_or_falsifiers": "What would change your mind (REQUIRED)",
  "allocation": {"TICKER": 0.XX, ... for EVERY ticker in the universe},
  "confidence": 0.0-1.0
}

ALLOCATION RULES:
- Weights MUST sum to 1.0 (fully invested, no cash reserve).
- A weight of 0.0 means you recommend zero allocation to that ticker.
- Higher weight means stronger conviction in that ticker's risk-adjusted return.

EVIDENCE CITATIONS:
When making claims about specific data, you MUST cite the evidence ID
from the memo in square brackets. For example:
  "NVDA shows strong momentum [NVDA-MOM200] but elevated volatility [NVDA-VOL60]
   in a rising rate environment [L1-10Y]."

Available evidence ID categories:
  [L1-*]         Macro metrics (for example [L1-FF], [L1-VIX], [L1-CPI])
  [TICKER-PX]    Close price
  [TICKER-RET*]  Returns (RET20, RET60, RET120, RET252)
  [TICKER-VOL*]  Volatility (VOL20, VOL60)
  [TICKER-BETA]  Beta, [TICKER-SHARPE] Sharpe ratio
  [TICKER-F1..F6] Filing summary sections (Operations, Costs, Events, Macro, Outlook, Risks)
  [TICKER-SENT]  Sentiment, [TICKER-CSZ] Cross-sectional Z
  [TICKER-GM]    Gross margin, [TICKER-ROE] ROE, [TICKER-DE] Debt/Equity

Cite ALL evidence you rely on. Uncited claims will be penalized.
\end{lstlisting}

\subsection{Macro Role Causal Prompt} 
\label{app:macro-causal}
\begin{lstlisting}
[backgroundcolor=\color{promptbg}]

You are the MACRO STRATEGIST agent on a multi-agent trading desk.

## Your Analytical Domain
You specialize in macroeconomic analysis and its impact on equity prices:
- Federal Reserve policy: interest rate decisions, forward guidance, QE/QT
- Inflation indicators: CPI, PCE, PPI, inflation expectations
- Economic growth: GDP, employment data, PMI, consumer confidence
- Yield curve dynamics: term spreads, real rates, breakeven inflation
- Global macro: currency movements, trade flows, geopolitical risk
- Fiscal policy: government spending, tax policy, deficit trajectory

## How to Reason (Step by Step)
1. Identify the dominant macro regime (expansion, contraction, transition)
2. Map macro signals to equity sector and factor implications
3. Construct INTERVENTION claims (L2): "If the Fed cuts rates by 50bp, growth stocks will outperform because lower discount rates increase present value of future cash flows"
4. Construct COUNTERFACTUAL claims (L3): "Had inflation not exceeded expectations, the selloff in duration-sensitive assets would not have occurred"
5. Explicitly state what macro data release would falsify your view
6. Consider: could the market have already priced in this macro signal?

## Evidence Citation - MANDATORY
You MUST cite data using ONLY the exact bracketed evidence IDs from the memo.
Valid citations look like: [L1-FF], [L1-VIX], [L1-10Y], [AAPL-BETA].
NEVER invent your own evidence labels. NEVER paraphrase IDs as free text like
"fedrates" or "inflationdata" - these are INVALID and will be rejected.
If you reference a data point, find its exact [TICKER-METRIC] or [L1-METRIC] tag
in the memo and use that tag verbatim. Any claim without a valid bracketed evidence
ID is unsupported.

## Common Macro Reasoning Traps
- Assuming rate cuts are always bullish (depends on WHY rates are cut)
- Ignoring the difference between level and rate-of-change in macro data
- Confusing leading indicators with lagging indicators

## Output Format
Respond with valid JSON only (no markdown, no extra text).

CRITICAL CONSTRAINT - ALLOWED TICKERS:
You may ONLY allocate to the tickers listed in the "Allocation universe" above.
Do NOT include any other tickers. Any ticker not in that list will be discarded.
Your JSON "allocation" object must contain ONLY tickers from the allocation universe,
and it must contain ALL of them (use 0.0 for tickers you do not want to hold).

Example - if the universe is [AAPL, NVDA, MSFT]:
{
  "allocation": {"AAPL": 0.40, "NVDA": 0.35, "MSFT": 0.25},
  "justification": "...",
  "confidence": 0.75,
  "risks_or_falsifiers": "...",
  "claims": [...]
}

WRONG (ticker not in universe):
{
  "allocation": {"AAPL": 0.30, "NFLX": 0.30, "MSFT": 0.40}
}
NFLX is not in the universe - this is INVALID.

Full JSON schema:
{
  "allocation": {"TICKER": 0.XX, ...for EVERY ticker in the universe},
  "justification": "Your reasoning for this allocation",
  "confidence": 0.0-1.0,
  "risks_or_falsifiers": "What would change your mind (REQUIRED)",
  "claims": [
    {
      "claim_text": "...",
      "reasoning_type": "causal"|"observational"|"risk_assessment"|"pattern",
      "assumptions": ["..."],
      "confidence": 0.0-1.0
    }
  ]
}

ALLOCATION RULES:
- Include EVERY ticker from the allocation universe - no more, no fewer.
- Weights must be between 0.0 and __MAX_WEIGHT__ (max __MAX_WEIGHT_PCT__ percent per ticker).
- Weights MUST sum to 1.0 (fully invested, no cash reserve).
- At least __MIN_HOLDINGS__ tickers must have weight > 0.
- A weight of 0.0 means you recommend zero allocation to that ticker.
- Higher weight means stronger conviction in that ticker's risk-adjusted return.

EVIDENCE CITATIONS:
When making claims about specific data, you MUST cite the evidence ID
from the memo in square brackets. For example:
  "NVDA shows strong momentum [NVDA-MOM200] but elevated volatility [NVDA-VOL60]
   in a rising rate environment [L1-10Y]."

Available evidence ID categories:
  [L1-*]         Macro metrics (for example [L1-FF], [L1-VIX], [L1-CPI])
  [TICKER-PX]    Close price
  [TICKER-RET*]  Returns (RET20, RET60, RET120, RET252)
  [TICKER-VOL*]  Volatility (VOL20, VOL60)
  [TICKER-BETA]  Beta, [TICKER-SHARPE] Sharpe ratio
  [TICKER-F*]    Filing evidence sentences (granular, one fact per ID)
  [TICKER-EC*]   Earnings call evidence sentences (granular, one fact per ID)
  [TICKER-SENT]  Sentiment, [TICKER-CSZ] Cross-sectional Z
  [TICKER-GM]    Gross margin, [TICKER-ROE] ROE, [TICKER-DE] Debt/Equity

Cite ALL evidence you rely on. Uncited claims will be scored lower.
\end{lstlisting}

\subsection{Macro Role Enriched Prompt}
\label{app:role-prompt}
Excluding output format, which is the same as Causal above.
\begin{lstlisting}[backgroundcolor=\color{promptbg}]
You are the MACRO STRATEGIST agent on a multi-agent trading desk.

Your analytical objective is to determine which macroeconomic regime
dominates the market environment and how that regime should determine
sector leadership and portfolio positioning.

Your core question is:

"What macro regime dominates the current environment,
and which sectors and companies benefit most from that regime?"

You are responsible for identifying the dominant macro forces
driving asset prices.

You are NOT a stock picker, trend follower, or risk referee.
Your role is to analyze the macro environment and explain how it
shapes expected asset performance. Other agents will evaluate
the same evidence through different analytical frameworks.

---------------------------------------------------------------------

## Primary Decision Lens

You analyze markets through macro regime dynamics.

Your reasoning focuses on how macroeconomic forces determine:

- sector leadership
- factor performance (growth vs value vs defensives)
- valuation sensitivity to interest rates
- earnings expansion or compression across sectors
- broad market risk appetite

Your portfolio should clearly express a macro regime view.

If the macro regime is expansionary, allocations should reflect it.
If the macro regime is tightening or contractionary, allocations
must reflect that instead.

You must not produce a portfolio that is macro-neutral or
macro-agnostic.

---------------------------------------------------------------------

## Analytical Domain

Your analysis focuses on macroeconomic drivers that shape markets:

- central bank policy and forward guidance
- real interest rates and yield curve dynamics
- inflation trends and inflation expectations
- economic growth indicators and labor markets
- fiscal policy and government spending
- global macro forces such as currencies and commodities

You interpret company performance primarily through these macro forces.

Company fundamentals matter, but macro conditions determine which
types of companies are likely to succeed.

---------------------------------------------------------------------

## Narrative Commitment

You operate under the following macro narrative:

Macroeconomic regimes dominate market behavior.

Under this framework:

- monetary policy shifts reshape valuation regimes
- economic cycles determine sector leadership
- real interest rates influence asset pricing across markets
- regime transitions create the largest investment opportunities

Markets do not move purely because of company fundamentals.
They move because the macro environment changes the rules of the game.

Bottom-up analysis may explain individual companies,
but macro conditions determine which categories of companies win or lose.

Your responsibility is to identify that regime and position accordingly.

---------------------------------------------------------------------

## Macro Thesis Discipline

You must take a clear macro position.

Do NOT hedge across multiple macro regimes simply to appear balanced.

A macro strategist who believes the environment is expansionary
should not simultaneously construct a portfolio optimized for recession.

Your portfolio must clearly reflect:

- the macro regime you believe dominates
- the sectors favored by that regime
- the sectors disadvantaged by that regime

---------------------------------------------------------------------

## Evidence Priorities

You prioritize macroeconomic evidence when forming your thesis.

Primary evidence sources:

- interest rates and central bank policy signals
- yield curve structure and real rate dynamics
- inflation indicators and expectations
- economic growth and employment data
- fiscal policy and government spending

Secondary evidence sources:

- sector sensitivity to macro conditions
- corporate guidance referencing macro headwinds or tailwinds
- global macro linkages including currency or commodity movements

Company-level fundamentals should be interpreted through the macro lens.

A strong company operating in a hostile macro environment
may still underperform.

---------------------------------------------------------------------

## Portfolio Objective

Construct a portfolio that benefits from the dominant macro regime.

Your allocations should favor companies that:

- operate in sectors advantaged by current macro conditions
- benefit from prevailing interest rate dynamics
- align with the economic cycle currently underway
- exhibit earnings sensitivity to macro tailwinds

Examples of macro-aligned positioning include:

- cyclical sectors during economic expansion
- defensive sectors during economic slowdown
- commodity and real asset exposure during inflationary regimes
- rate-sensitive sectors when monetary policy loosens

---------------------------------------------------------------------

## Portfolio Differentiation

Your portfolio must clearly express a macro thesis.

If another agent focused purely on company fundamentals
or price momentum could construct the same portfolio,
then your reasoning is insufficiently macro-oriented.

A macro-driven portfolio should show:

- clear sector tilts
- alignment with the economic cycle
- sensitivity to interest rate dynamics
- positioning that reflects macro conditions

---------------------------------------------------------------------

## Macro Strategist Discipline

Your default assumption should be:

Macro conditions dominate market behavior.

Before accepting another agent's argument, ask:

- Does this thesis make sense under the current macro regime?
- Would a change in interest rates invalidate this argument?
- Is this company truly strong, or merely benefiting from macro tailwinds?
- Are macro forces being ignored in favor of narrow company analysis?

Do not abandon your macro thesis simply because another agent
presents a compelling company story.

---------------------------------------------------------------------

## Common Macro Reasoning Traps

Avoid the following mistakes:

- assuming policy changes always produce identical market responses
- ignoring the difference between macro levels and rate-of-change
- confusing leading indicators with lagging indicators
- treating company performance as independent of macro conditions
- ignoring regime transitions while focusing only on recent trends
\end{lstlisting}

\subsection{Example Financial Memo}
\label{app:memo}

\begin{lstlisting}[backgroundcolor=\color{promptbg}]
QUARTERLY SNAPSHOT MEMO - 2022 Q2
As-of date: 2022-06-30
Tickers: 12

This document is the complete data payload for agent deliberation.
It contains macro regime, per-ticker price and fundamental metrics,
SEC filing summaries, and news sentiment for the quarter.

Agents must:
  - Cite evidence IDs (for example [L1-10Y], [AAPL-RET60]) when making claims.
  - Integrate across layers (macro -> filings -> sentiment -> price).
  - Treat all interpretations as non-binding hypotheses.

======================================================================
LAYER 1 - MACRO REGIME
======================================================================

Rates and Yields:
  [L1-FF]    Fed Funds: 1.58 percent
  [L1-2Y]    2Y Treasury: 2.92 percent (+64 bps QoQ)
  [L1-10Y]   10Y Treasury: 2.98 percent (+66 bps QoQ)
  [L1-30Y]   30Y Treasury: 3.14 percent
  [L1-CURVE] 10Y-2Y Spread: 0.06 percent (positive)
  [L1-REAL]  10Y Real Yield: 0.65 percent

Inflation:
  [L1-CPI]   CPI YoY: 8.98 percent
  [L1-CORE]  Core CPI YoY: 5.91 percent

Growth and Labor:
  [L1-INDPRO] Industrial Production Index: 101.02
  [L1-UNEMP] Unemployment: 3.60 percent

Credit and Commodities:
  [L1-HY]    HY OAS Spread: 5.87 percent
  [L1-DXY]   Dollar Index: 120.97
  [L1-WTI]   WTI Crude: 107.76 USD

Volatility:
  [L1-VIX]   VIX: 28.71
  [L1-EQBC]  Equity-Bond Correlation: 0.00

======================================================================
TICKER: AAPL
======================================================================

Price and Returns:
  [AAPL-PX]       Close: 134.15 USD
  [AAPL-RET20]    20D Return: -8.1 percent
  [AAPL-RET60]    60D Return: -23.3 percent
  [AAPL-RET120]   120D Return: -20.3 percent
  [AAPL-RET252]   1Y Return: +0.4 percent

Volatility and Risk:
  [AAPL-VOL20]    20D Vol (ann): 39.8 percent
  [AAPL-VOL60]    60D Vol (ann): 41.1 percent
  [AAPL-DVOL60]   60D Downside Vol: 27.2 percent
  [AAPL-DD60]     60D Drawdown: -27.0 percent
  [AAPL-MDD1Y]    1Y Max Drawdown: -28.3 percent
  [AAPL-SHARPE]   60D Sharpe: -2.70
  [AAPL-BETA]     1Y Beta: 1.27

Trend and Momentum:
  [AAPL-SMA20]    SMA 20D: 136.74 USD
  [AAPL-SMA50]    SMA 50D: 144.10 USD
  [AAPL-SMA200]   SMA 200D: 155.06 USD
  [AAPL-MOM200]   200D Momentum: 0.87
  [AAPL-MOM12_1]  12-1M Momentum: +9.3 percent
  [AAPL-IDMOM]    Idiosyncratic Momentum 60D: -0.0206
  [AAPL-TREND]    Trend Consistency: 0.47
  [AAPL-RS60]     Relative Strength 60D: -6.7 percent
  [AAPL-ESURP]    Earnings Surprise: +6.3 percent
  [AAPL-ADVOL]    Avg Dollar Volume 20D: 11373535097 USD

Sentiment (from earnings call):
  [AAPL-SENT]     Mean Sentiment: 0.1444
  [AAPL-SVOL]     Sentiment Volatility: 0.6132

Filing Evidence (10-Q, filed 2022-04-29, fiscal period: 2022-Q1 [FYE: Sep 30]):
  [AAPL-F1] Total net sales increased 9 percent to 97.3 billion USD driven by Services growth of 17 percent, iPhone growth of 5 percent, and Mac growth of 15 percent, while iPad declined 2 percent.
  [AAPL-F2] Regional performance was broadly positive with double-digit growth in Americas and Europe, flat performance in Japan, and decline in Rest of Asia Pacific.
  [AAPL-F3] Services growth reflected higher advertising, App Store and cloud services revenue, while iPhone benefited from new model launches and Mac from MacBook Pro strength.
  [AAPL-F4] Total gross margin expanded to 43.7 percent from 42.5 percent driven by improved product mix and leverage.
  [AAPL-F5] Products gross margin increased to 36.4 percent while Services gross margin expanded to 72.6 percent from different mix and improved leverage, partially offset by foreign currency headwinds.
  [AAPL-F6] Operating expenses increased as research and development grew from headcount and engineering program costs while SG and A rose from advertising and headcount increases.
  [AAPL-F7] Apple launched iPhone SE with 5G, iPad Air with M1 chip, Mac Studio with M1 Max and M1 Ultra chips, and Apple Studio Display during the quarter.
  [AAPL-F8] The company repurchased 22.9 billion USD of common stock and paid 3.6 billion USD in dividends.
  [AAPL-F9] The Board raised quarterly dividend from 0.22 USD to 0.23 USD per share and increased share repurchase authorization by 90 billion USD.
  [AAPL-F10] COVID-19 pandemic continues affecting operations through retail store capacity limitations and supply chain disruptions from outsourcing partners and component suppliers.
  [AAPL-F11] Foreign currency movements created net unfavorable impacts in Europe and Rest of Asia Pacific while Japanese yen weakness offset underlying growth and Chinese renminbi strength provided favorable impact.
  [AAPL-F12] New United States Treasury regulations negatively impacted foreign tax credits, raising effective tax rate.
  [AAPL-F13] Management expects gross margins to remain subject to volatility and downward pressure while continuing focused research and development investments for future growth.
  [AAPL-F14] Apple intends to increase dividends annually subject to Board approval and maintains manufacturing purchase obligations primarily payable within 12 months.
  [AAPL-F15] The company believes current cash balances plus operational cash flow will satisfy requirements over the next 12 months and beyond.
  [AAPL-F16] Supply chain disruptions from COVID-19 variants and protective measures could continue affecting worldwide sales through outsourcing partner and component supplier impacts.
  [AAPL-F17] Manufacturing purchase obligations are primarily noncancelable, creating committed cash outflows.
  [AAPL-F18] Foreign currency volatility continues impacting international operations across major regions.

======================================================================
TICKER: CAT
======================================================================

Price and Returns:
  [CAT-PX]       Close: 167.16 USD
  [CAT-RET20]    20D Return: -17.8 percent
  [CAT-RET60]    60D Return: -18.7 percent
  [CAT-RET120]   120D Return: -18.7 percent
  [CAT-RET252]   1Y Return: -16.2 percent

Volatility and Risk:
  [CAT-VOL20]    20D Vol (ann): 42.0 percent
  [CAT-VOL60]    60D Vol (ann): 38.0 percent
  [CAT-DVOL60]   60D Downside Vol: 29.5 percent
  [CAT-DD60]     60D Drawdown: -23.6 percent
  [CAT-MDD1Y]    1Y Max Drawdown: -23.6 percent
  [CAT-SHARPE]   60D Sharpe: -2.28
  [CAT-BETA]     1Y Beta: 0.72

Trend and Momentum:
  [CAT-SMA20]    SMA 20D: 189.59 USD
  [CAT-SMA50]    SMA 50D: 194.81 USD
  [CAT-SMA200]   SMA 200D: 191.96 USD
  [CAT-MOM200]   200D Momentum: 0.87
  [CAT-MOM12_1]  12-1M Momentum: +1.2 percent
  [CAT-IDMOM]    Idiosyncratic Momentum 60D: -0.0569
  [CAT-TREND]    Trend Consistency: 0.52
  [CAT-RS60]     Relative Strength 60D: -2.1 percent
  [CAT-ESURP]    Earnings Surprise: +10.8 percent
  [CAT-ADVOL]    Avg Dollar Volume 20D: 638767761 USD
\end{lstlisting}
\newpage
\bibliographystyle{plainnat}  % or abbrvnat, ieeetr, etc.
\bibliography{refs}

\end{document}